\documentclass[ aip,amsmath,amssymb,reprint]{revtex4-1}
\newcommand{\vc}{\mathbf}
\usepackage{color}
\usepackage{graphicx}
\usepackage{dcolumn}
\usepackage{bm}
\usepackage{appendix}

\usepackage[utf8]{inputenc}
\usepackage[T1]{fontenc}
\usepackage{mathptmx}

\begin{document}

\preprint{AIP/123-QED}

\title{Effects of Coulomb Coupling On Friction In Strongly Magnetized Plasmas}

\author{David J. Bernstein}
\affiliation{Department of Physics and Astronomy, University of Iowa, Iowa City, Iowa 52242, USA}
\author{Scott D.~Baalrud}
\email{baalrud@umich.edu}
\affiliation{Department of Nuclear Engineering and Radiological Sciences, University of Michigan, Ann Arbor, MI 48109, USA}

\date{\today}

\begin{abstract}
The friction force on a test particle traveling through a plasma that is both strongly coupled and strongly magnetized is studied using molecular dynamics simulations. 
In addition to the usual stopping power component aligned antiparallel to the velocity, a transverse component that is perpendicular to both the velocity and Lorentz force is observed. 
This component, which was recently discovered in weakly coupled plasmas, is found to increase in both absolute and relative magnitude in the strongly coupled regime. 
Strong coupling is also observed to induce a third component of the friction force in the direction of the Lorentz force.
These first-principles simulations reveal novel physics associated with collisions in strongly coupled, strongly magnetized, plasmas that are not predicted by existing kinetic theories. 
The effect is expected to influence macroscopic transport in a number of laboratory experiments and astrophysical plasmas. 

\end{abstract}

\maketitle

\section{Introduction}
\label{sec:intro}
Both natural and laboratory plasmas often occur in the presence of external magnetic fields.
In most instances, the magnetic field only weakly magnetizes the plasma in the sense that the particle gyrofrequency $\omega_c \equiv qB/cm$ (where $q$ and $m$ are the particle charge and mass, $B$ the magnetic field strength, and $c$ the speed of light) is much smaller than the plasma frequency $\omega_p \equiv \sqrt{4 \pi n q^2 / m}$ (where $n$ is the number density) \cite{Baalrud_Daligault_MagPhases}. 
The ordering $\beta \equiv \omega_c/\omega_p \ll 1$ is used as an expansion parameter in traditional plasma kinetic theory, leading to the result that the magnetic field does not influence microscopic physics at the scale of collisions.
It is interesting to explore how the fundamental physics of transport changes when a plasma is strongly magnetized ($\beta > 1$).
For example, recent work has shown that the friction force on a test particle, which describes the most basic form of momentum transport, is fundamentally altered by strong magnetization~\cite{Lafleur_Baalrud_2019,Jose_Baalrud_2020,bldb_2020,Lafleur_Baalrud_2020}.
Besides being an interesting regime to study from a basic physics perspective, plasmas in many experiments and in nature are strongly magnetized.
These include experiments on antimatter traps, \cite{Surko_Fajan,antimatter_2015,antimatter_2004}, nonneutral plasmas, \cite{nonneutral_PRL_1977,nonneutral_PRL_1980,nonneutral_PRL_1980,nonneutral_PhysFluids_1980}, and ultracold neutral plasmas \cite{Ultracold_2,Killian_UCNP}, as well as natural systems such as neutron star atmospheres \cite{neutron_star_1}.
In addition to being strongly magnetized, the plasmas in these systems can reach regimes of strong Coulomb coupling, i.e. when the average inter-particle potential energy exceeds the average kinetic energy per particle \cite{Ichimaru}.
Here, we evaluate the combined influence of strong magnetization and strong coupling on the friction force using first-principles molecular dynamics (MD) simulations.

\begin{figure}
\centering
\includegraphics[width=5cm]{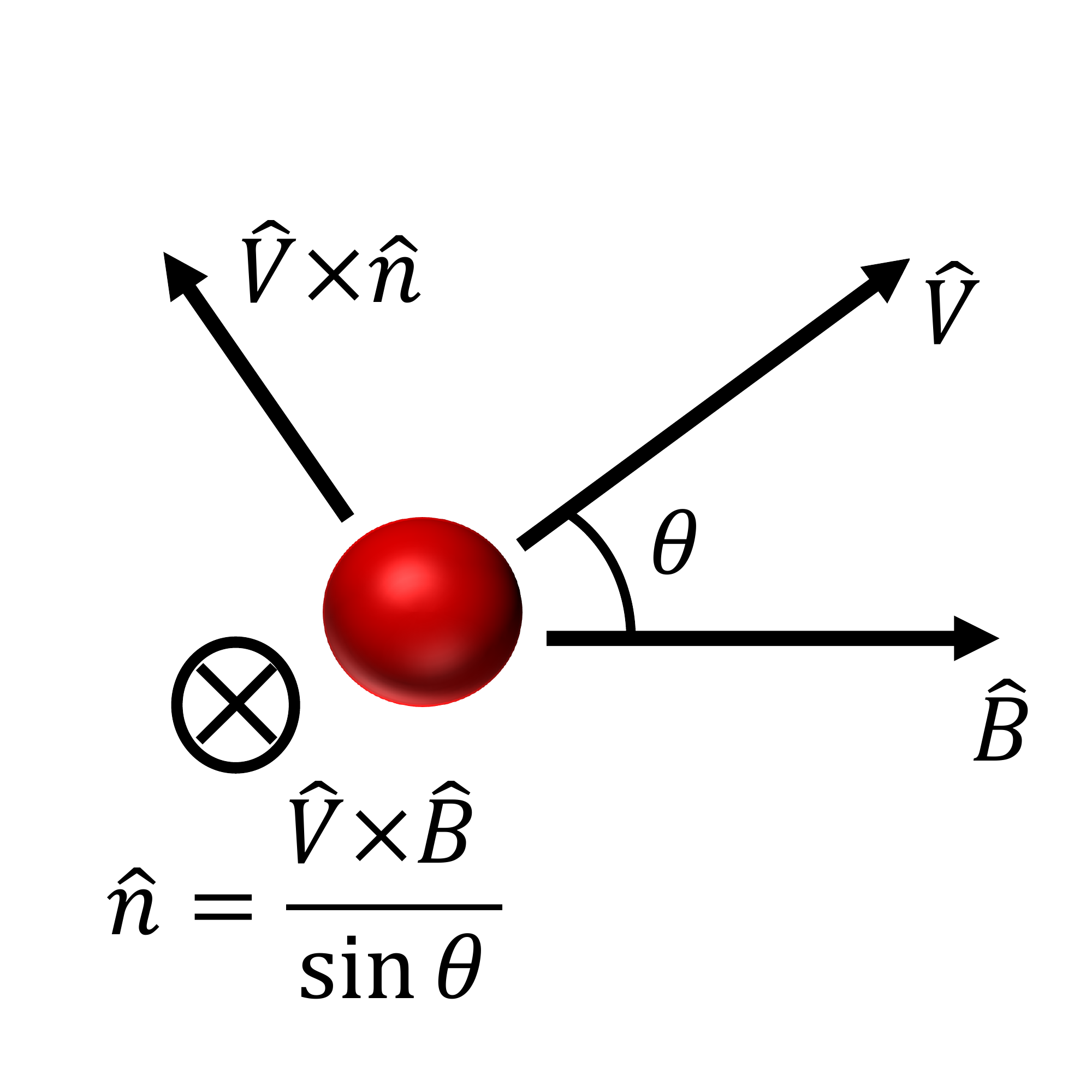}
\caption{Coordinates of the test-particle (red circle) velocity $\vc{V}$ and magnetic field $\vc{B}$. 
The Lorentz force direction $\hat{\vc{n}}$ points into the page.
}
\label{fig:geometry}
\end{figure}

The average motion of a test particle traveling through a plasma on timescales long compared to the collision time can be approximated as
\begin{equation}
\label{eq:eom}
M \frac{d \vc{V}}{dt} = \frac{Q}{c} \vc{V} \times \vc{B} + \vc{F},
\end{equation}
where $M$ is the mass of the test particle, $Q$ the charge, $\vc{V}$ the velocity, $\vc{B}$ the external magnetic field, and $\vc{F}$ the friction force due to drag from the background plasma.
The friction force in a weakly coupled and weakly magnetized plasma acts antiparallel to the test particle's velocity $\vc{F} = F_v \hat{\vc{V}}$ where $\hat{\vc{V}} = \vc{V}/V$, and is commonly referred to as stopping power~\cite{Zwick_Review,Mag_dEdx_Book}.
Recent results have shown a surprising effect that strong magnetization causes the friction force to also have a transverse component that acts perpendicular to the Lorentz force and test particle velocity \cite{Lafleur_Baalrud_2019,Jose_Baalrud_2020,Lafleur_Baalrud_2020,bldb_2020}
\begin{equation}
\label{eq:fric_mag}
\vc{F} = F_v \hat{\vc{V}} + F_\times \hat{\vc{V}} \times \hat{\vc{n}},
\end{equation}
where $F_\times$ is the transverse component, and $\hat{\vc{n}} = \hat{\vc{V}} \times \hat{\vc{B}}/\sin \theta$ is the unit vector of the Lorentz force where $\hat{\vc{B}} = \vc{B}/B$, and $\theta$ is the angle between $\vc{V}$ and $\vc{B}$ in the plane defined by the two vectors; see Fig.~\ref{fig:geometry}. 
The existence of this transverse force was first predicted using linear response theory,~\cite{Lafleur_Baalrud_2019} and was later confirmed using MD simulations~\cite{bldb_2020}. 
It has also recently been modeled using a new collisional kinetic theory for strongly magnetized plasmas~\cite{Jose_Baalrud_2020}.
Since the transverse friction transfers momentum between the directions parallel and perpendicular to the magnetic field, it significantly alters particle dynamics, as well as macroscopic transport~\cite{Lafleur_Baalrud_2019}.
These previous studies concentrated on the weakly coupled regime [$\Gamma \ll 1$ in Eq.~(\ref{eq:Gamma})]. 
However, many strongly magnetized plasmas, such as those in the previously mentioned examples, are also strongly coupled. 
Here, we extend this investigation into the strongly coupled regime ($\Gamma >1$).

The Coulomb coupling strength in a one-component plasma (OCP) is quantified by the Coulomb coupling parameter 
\begin{equation}
\label{eq:Gamma}
    \Gamma \equiv \frac{q^2/a}{k_BT}
\end{equation}
where $a=(3/4 \pi n) ^{1/3}$ is the average inter-particle spacing, $k_B$ is the Boltzmann constant, and $T$ the plasma temperature \cite{Ichimaru,BausHansen}.
Considering the friction force on a massive test particle, the influence of strong coupling has been studied in unmagnetized plasmas using both theory \cite{Zwick_Review,Peter_Meyer-ter-Vehn,Petrasso,EPT_Conf} and MD simulations \cite{Zwick_Review,bbd_2019,MD_Grabowski}.
These show that strong coupling causes the Bragg peak to shift to a higher speed relative to the thermal speed of the background plasma, and for the stopping power curve to broaden~\cite{bbd_2019}.
It is unknown how strong magnetization influences these results. 
Previous MD simulations verify the existence of the transverse force in plasmas with $\Gamma = 0.1$ and 1, but have not calculated the friction when $\Gamma > 1$ \cite{bldb_2020}.
Is the transverse friction ($F_\times$) present in the strongly magnetized regime?
If so, how does strong coupling influence it? 
Furthermore, can the friction force be characterized by only two vector components in this regime, or is a third component also required?

Our MD simulations show that the transverse friction increases in both absolute and relative magnitude in the strongly coupled regime, and its dependence on the test particle's speed qualitatively changes. 
For instance, unlike in weakly coupled plasmas, $F_\times$ does not change sign depending on the speed of the test particle.
Moreover, the friction is found to not lie in the plane defined by Eq.~(\ref{eq:fric_mag}), but to also depend on a third component ($F_n$) oriented along the same direction as the Lorentz force
\begin{equation}
\label{eq:friction}
\vc{F} = F_v \hat{\vc{V}} + F_\times \hat{\vc{V}} \times \hat{\vc{n}}+F_n \hat{\vc{n}}.
\end{equation} 
The $F_n$ component is not present at the $\beta$ values investigated when $\Gamma \ll 1$ \cite{Lafleur_Baalrud_2019,Jose_Baalrud_2020}.
Depending on the speed of the test particle, $F_n$ acts either parallel or antiparallel to the Lorentz force.
As coupling increases, both the $F_\times$ and $F_n$ components increase in magnitude compared to the $F_v$ component.
As the coupling increases, the absolute magnitude of each component increases, the peak force shifts to higher a test particle speed, and the force curve broadens as a function of test particle speed.
These results demonstrate qualitatively new physics features associated with the friction on a test particle. 
In turn, they are expected to translate to qualitatively new features in macroscopic transport, such as electrical conductivity \cite{Braginskii}.

With no theory applicable under the conditions of strong coupling and strong magnetization, these first-principles MD simulations are a useful tool \cite{MDbook}.
Because the assumptions underlying the simulations are minimal (classical Coulomb interactions), they provide a first-principles method to explore new regimes. 
The data obtained is expected to provide a benchmark for future theories.

\section{Simulation Setup and Analysis}
\label{sec:sims}

The dynamics of test particles traveling through the magnetized OCP were calculated using the MD code described in Ref.~\onlinecite{Code}. 
The OCP consists of single species of particles with mass $m$ and charge $q$ with an inert neutralizing background~\cite{BausHansen}.
The magnetized OCP is fully parameterized by $\beta$ and $\Gamma$~\cite{BausHansen,Ott_Bonitz_PRL}.
In this model, the test particle's mass is quantified by the ratio of its mass and that of the background particles, $M/m$, and its speed relative to the thermal speed of the background, $V/v_T$, where $v_T = \sqrt{2k_BT/m}$. 
Although simplified, this model provides an accurate representation of friction in most real plasmas because the friction force is predominately determined by the species with a thermal speed close to the speed of the test charge~\cite{Lafleur_Baalrud_2020}. 
Because it can be parameterized by only $\Gamma$ and $\beta$, it is also an ideal system to isolate the effects of strong coupling and strong magnetization.

All particles were taken to interact via the Coulomb force.
For numerical efficiency, this was modeled using the Ewald-summation technique, which splits the force into short and long-range components \cite{MDbook}. 
This was implemented using the particle-particle-particle-mesh algorithm \cite{MDbook}.
Periodic boundaries about the cubic simulation domain were used to simulate an infinite plasma.
Convergence was obtained so that the computed friction force was independent of the domain size. 
Depending on $\Gamma$, either $N = 5 \times 10^4$ or $1 \times 10^4$ particles were sufficient to ensure that this conditions was met; see Tbl.~\ref{tab:tab1}.

\begin{table}
\begin{tabular}{ccccccccc}
\hline
\hline
$\Gamma$ &$N$ & $L$ & $\beta$ &$V_0$ & $\theta$&\\
\hline
0.1& $ 5 \times 10^4 $ & 59.386$a$ &10  & 0-3$v_T$&$22.5^{\circ}$\\
1  & $1 \times 10^4 $  & 34.729$a$ &0, 1, 10 & 0-3$v_T$&$0^{\circ}$, $22.5^{\circ}$, 90, 157.5,-90\\
10 & $1 \times 10^4 $ & 34.729$a$ &0, 1, 10  &0-10$v_T$ & $22.5^{\circ}$\\
100 & $1 \times 10^4 $ & 34.729$a$ &10 &0-25$v_T$ & $22.5^{\circ}$\\
\hline
\hline
 \end{tabular}
\caption{Simulation inputs: Coulomb coupling parameter $\Gamma$, number of particles $N$, length $L$ of the simulation unit-cell, test particle speed $V_0$, magnetization parameter $\beta$, and angle $\theta$ of the test particle velocity with respect to the magnetic field.}
\label{tab:tab1}
\end{table}

All simulations started by equilibrating an unmagnetized OCP at a fixed $\Gamma$ for $500 \omega_p^{-1}$ with a velocity scaling thermostat \cite{MDbook}.
This provided enough time to remove any effects of the initial random placement of particles, and allowed the system to reach equilibrium at the chosen $\Gamma$.
According to the Bohr–van Leeuwen theorem, the equilibrated state is the same with or without a magnetic field \cite{Pathria}.
The magnetic field was not included during the equilibration stage so that the relaxation to equilibrium was faster. 
The magnetic field was turned on after the equilibration, when the test particle was introduced.
Time was discretized into timesteps of $0.001\omega_p^{-1}$, which was small enough to resolve collisions and the gyration of particles over the range of $\Gamma$ and $\beta$ values investigated.
After the initial $500 \omega_p^{-1}$ equilibration stage, a large configuration of statistically-independent initial conditions were obtained by extending the equilibration stage for another $30,000 \omega_p^{-1}$ and saving the particle positions and velocities at every $1 \omega_p^{-1}$. 

\begin{figure}
\centering
\includegraphics[width=8.5cm]{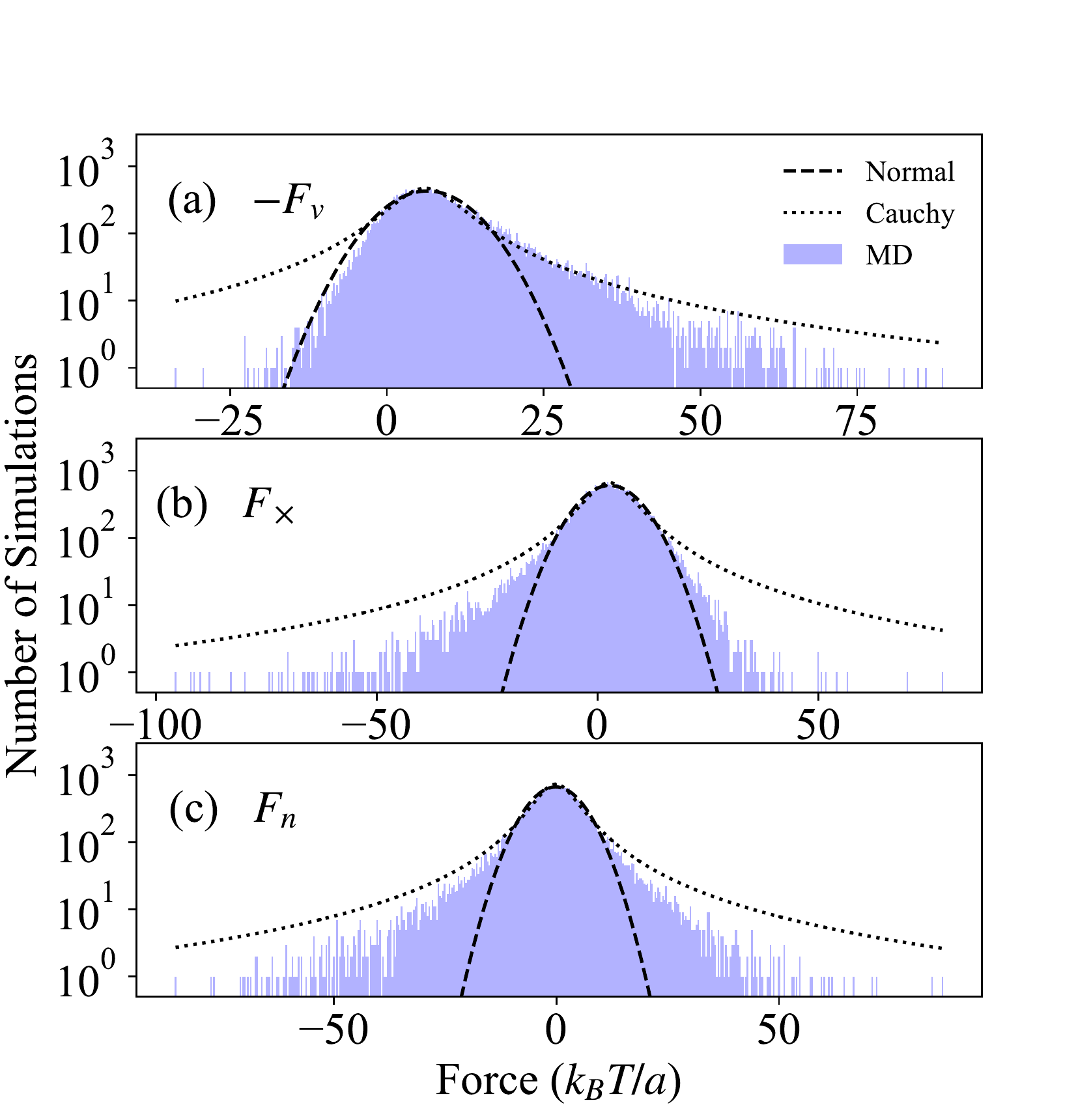}
\caption{Histogram of the friction force  components [$-F_v$ in (a), $F_\times$ in (b), and $F_n$ in (c)] computed from each of the 30,000 simulations from a simulation with $\Gamma = 10$, $\beta = 10$ and $\theta = 22.5^\circ$ and a test-particle speed of $3v_T$. 
Best fit lines to a normal distribution (dashed line), and Cauchy distribution (dotted line) are also shown. 
Each histogram consists of 500 bins.}

\label{fig:hists}
\end{figure}

\begin{figure*}
\centering
\includegraphics[width=18cm]{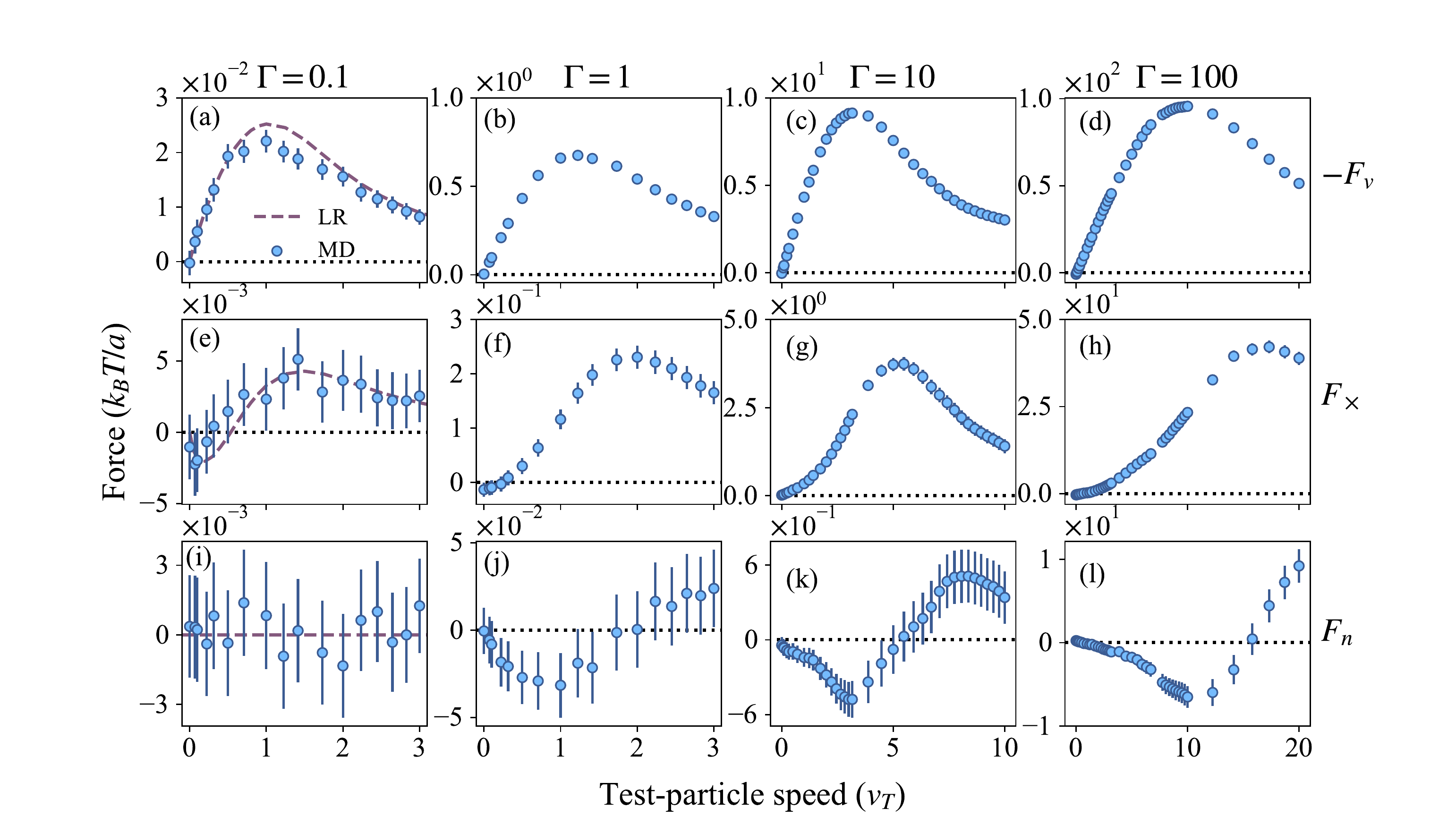}
\caption{Friction force (units of $k_BT/a$) as a function of test-particle speed (units of $v_T$) for $\beta = 10$, $\theta = 22.5^\circ$, and $\Gamma = 0.1$ [panels (a), (e), and (i)], 1 [panels (b), (f), and (j)], 10 [panels (c), (g), and (k)], and 100 [panels (d), (h), and (l)].
The stopping power components $-F_v$ are shown in panels (a), (b), (c) , and (d), the transverse components $F_\times$ in panels (e), (f), (g), and (h), and component in the direction of the Lorentz force $F_n$ in panels (i), (j), (k), and (l).
Predictions from a linear response theory~\cite{Lafleur_Baalrud_2019} are included as a purple dashed line with the $\Gamma = 0.1$ data.
The $\Gamma = 0.1$ and 1 results were presented in Ref.~\onlinecite{bldb_2020}.
}
\label{fig:force}
\end{figure*}

For each of the 30,000 initial configurations, a friction force calculation was conducted.
Each calculation started by turning off the thermostat and turning on an external magnetic field oriented along the $z-$direction of a Cartesian coordinate system with a strength corresponding to $\beta = 0$, 1, or 10.
A massive unmagnetized test particle (mass $M=1000m$ and charge $Q=q$) was then placed in the simulation domain and launched at an angle $\theta$ (with respect to the magnetic field in the $x-z$ plane) with initial speed $V_0$ (the test particle was not present during the equilibration stage).
The test particle momentum was fixed for the first $2 \omega_p^{-1}$ in order to remove transient effects from the abrupt insertion of the test particle in the plasma.
After this short period, the test particle was then free to interact with the plasma; its momentum was no longer fixed.
The force on the test particle in the $x$, $y$, and $z$ directions were recorded every 10 timesteps (every $0.01 \omega_p^{-1}$) for $1 \omega_p^{-1}$ yielding a time series of the force.
Because the test particle is massive, the approximation that it is unmagnetized over the short simulation duration is valid for the range of parameters investigated ($\beta \leq 10$).
Likewise, the $1 \omega_p^{-1}$ time of the data collection stage is expected to be short enough to represent an instantaneous force on the massive test particle. 

After all 30,000 simulations concluded for a given $\beta$, $V_0$, and $\theta$, the average force was computed in two steps.
First, each of the $1\omega_p^{-1}$ time series were averaged to give a single value for the instantaneous force associated with each of the 30,000 independent time series. 
These were recorded in the Cartesian domain, and then converted to the $\hat{\vc{V}}$-$\hat{\vc{B}}$-$\hat{\vc{n}}$ coordinate system (Fig.~\ref{fig:geometry}) using
\begin{subequations}
\begin{align}
F_v &= F_x \sin \theta + F_z \cos \theta \\ 
F_\times &= F_x \cos \theta - F_z \sin \theta \\
F_{n} &= -F_y.
\end{align}
\end{subequations}
Second, the 30,000 values were averaged to provide a single value for the instantaneous friction force. 

The large number of simulations was necessary to reduce noise \cite{bldb_2020}.
As shown in Fig.~\ref{fig:hists}, the distributions of the forces have fat tails and are highly skewed.
The nature of the statistics of these distributions is not known. 
In Fig.~\ref{fig:hists}, a best fit normal distribution and best fit Cauchy distribution are shown.
The bulk of the distribution is well approximated by the normal distribution, but the tails are better approximated by the Cauchy distributions (although not shown, the velocities are well approximated by $\kappa$-distributions \cite{kappa}).
However, none of these forms account for skew, which is evident in all three components of the force vector.
Despite these skewed and fat tailed distributions with unknown analytic forms, the standard deviation of the mean $\sigma_m=\sigma/\sqrt{N}$, where $\sigma$ is the standard deviation of the data and $N=30,000$ is the number of simulations, provides a good statistic for quantifying the error of the mean forces per the central limit theorem~\cite{Stats_book}.
All error bars were computed from $\pm  2.576 \sigma_m$, which corresponds to $99 \%$ confidence.

\section{Results}
\label{sec:results}
\subsection{Influence of coupling strength} 

Figure~\ref{fig:force} shows how Coulomb coupling influences the friction force in strongly magnetized plasmas. 
This data spans weak coupling ($\Gamma = 0.1$), moderate coupling ($\Gamma = 1$), and strong coupling ($\Gamma = 10$ and 100) regimes. 
Here, the magnetization strength is $\beta=10$, and the angle between the velocity and magnetic field is $\theta = 22.5^\circ$. 
Results at weak coupling are compared with the predictions of a linear response theory from Ref.~\onlinecite{Lafleur_Baalrud_2019}. 
The good agreement between theory and MD simulations at these conditions was previously reported in Ref.~\onlinecite{bldb_2020}. 
The new data at higher $\Gamma$ values shows that the trends of the stopping power component ($-F_v$) are qualitatively similar to what has been observed in unmagnetized plasmas [Fig.~\ref{fig:force}(a), (b), (c), and (d)]; the curve broadens with increasing $\Gamma$ and the peak stopping power (Bragg peak) increases in units of $k_BT/a$ and shifts to a higher speed \cite{bbd_2019}. 

The transverse force ($F_\times$) is found to be non-negligible throughout the range of $\Gamma$ values [Fig.~\ref{fig:force}(e), (f), (g), and (h)]. 
In fact, it is found to increase in absolute magnitude (in units of $k_BT/a$), as well as its magnitude in comparison to the stopping component, as $\Gamma$ increases. 
Unlike the stopping component, the transverse component has some qualitative differences in the strongly coupled regime.
In particular, the sign change that is observed at low speeds in the weakly coupled regime is not observed at moderate or strong coupling. 
A positive sign of the transverse force corresponds to a force component that acts to increase the gyroradius of the test particle, as described in Ref.~\onlinecite{Lafleur_Baalrud_2019}. 
As with the stopping component, the peak transverse component is found to shift to higher speed at stronger coupling, and the curve to broaden. 
These results show that the qualitative effect predicted by linear response theory in the weakly coupled regime \cite{Lafleur_Baalrud_2019} extends into the strongly coupled regime, where that theory does not apply. 

The most surprising feature of these results is that there is a component of the friction force in the direction of the Lorentz force ($F_n$) [Fig.~\ref{fig:force}(j), (k), and (l)]. 
This component is not present at weak coupling [$\Gamma = 0.1$ in  Fig.~\ref{fig:force}(i)] when $\beta =10$, and is only slightly greater than the noise at moderate coupling [$\Gamma = 1$ in Fig.~\ref{fig:force}(j)], but is easily computed far above the noise level at strong coupling [$\Gamma = 10$ and 100 in Figs.~\ref{fig:force}(k) and (l), respectively]. 
It is observed to change sign depending on the test particle speed.
A positive sign of $F_n$ corresponds to a force that increases the gyrofrequency of the test particle, while a negative sign acts to decrease the gyrofrequency. 
A previous theory that first predicted the transverse force was based on a linear response approach that applies only at weak coupling ($\Gamma \ll 1$) \cite{Lafleur_Baalrud_2019}.
That theory predicts $F_n=0$ as a basic symmetry property of the underlying linear response function. 
The data shown in Fig.~\ref{fig:force} show that this symmetry is broken in the strongly coupled regime. 
Linear response theory assumes that interactions between particles are well represented by only weak long-range interactions. 
The breakdown of this prediction at strong coupling implies that the $F_n$ component of the friction force is associated with strong short-range interactions that are excluded in the linear response approach. 

\begin{figure}
\centering
\includegraphics[width=8.5cm]{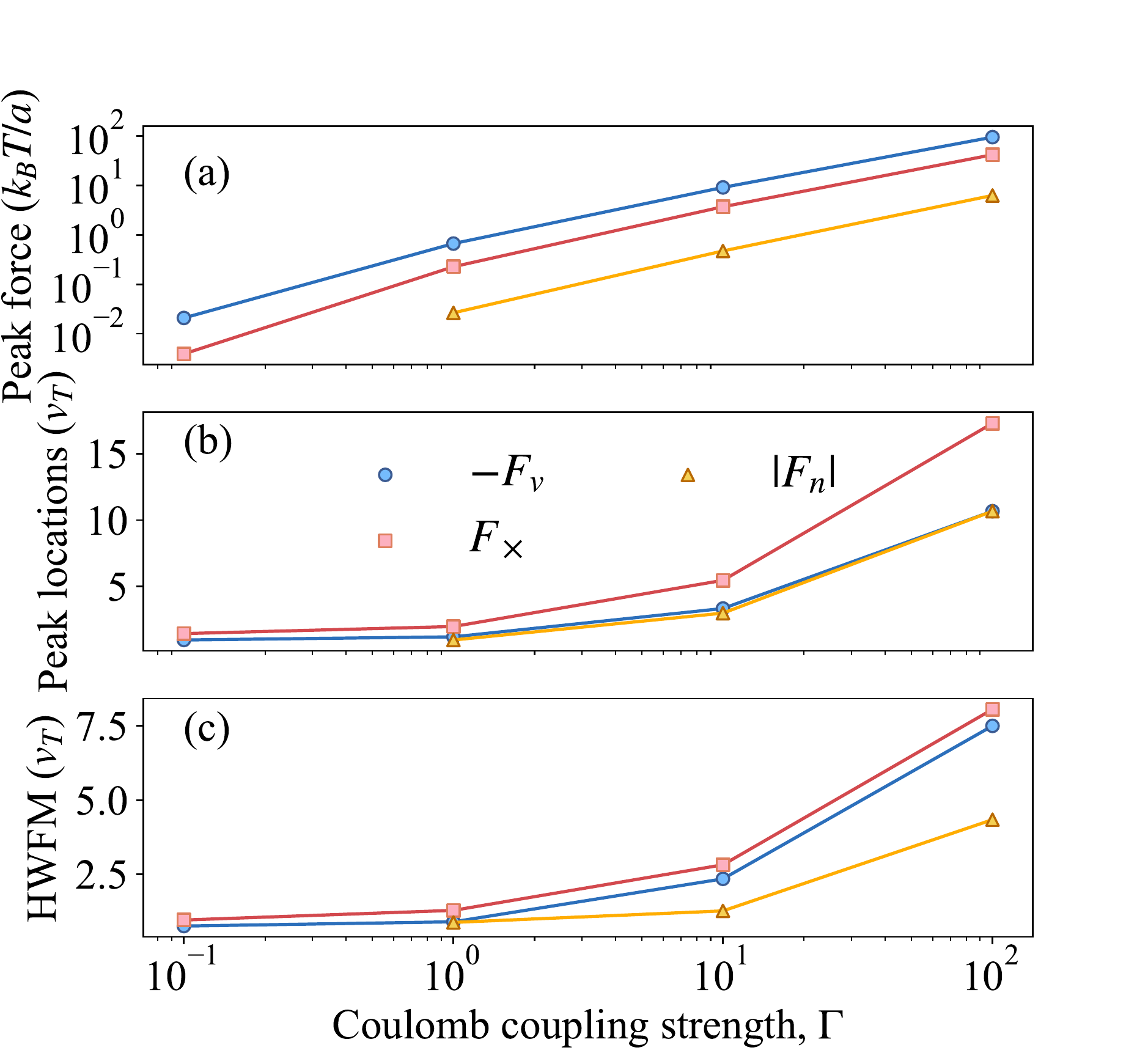}
\caption{
(a) Magnitude of the first extremum of each component of the friction force (for $F_n$, this is taken from the first minimum). 
(b) Speed at which the extrema occur (peak locations).
(c) Half-width full-maximum for the curve associated with each component of the friction force. 
This corresponds to the data set from Fig.~\ref{fig:force}; $\beta =10$, $\theta = 22.5^\circ$.
}
\label{fig:trends}
\end{figure}

Figure~\ref{fig:trends} shows trends of qualitative features of the force component curves as the coupling strength varies. 
As the coupling strength increases the magnitudes of each component increases [Fig.~\ref{fig:trends}(a)], the speed at which the peak force occurs increases [Fig.~\ref{fig:trends}(b)], and the curve associated with each component broadens [Fig.~\ref{fig:trends}(c)].
In Fig.~\ref{fig:trends}(c), the half-width at full maximum was calculated by recording the speed at which the force is half of the respective peak force value (from low to high speeds), then calculating the difference between the velocity at which the peak force occurs and the velocity at which the force is half of the peak value. 
These basic trends were observed in the stopping component in previous simulations for the unmagnetized OCP~\cite{bbd_2019,MD_Grabowski}.
As seen in Fig.~\ref{fig:trends}, the effects of strong coupling carry over to all components of the friction when $\beta > 1$.

\subsection{Influence of magnetization strength}

\begin{figure}
\centering
\includegraphics[width=8.5cm]{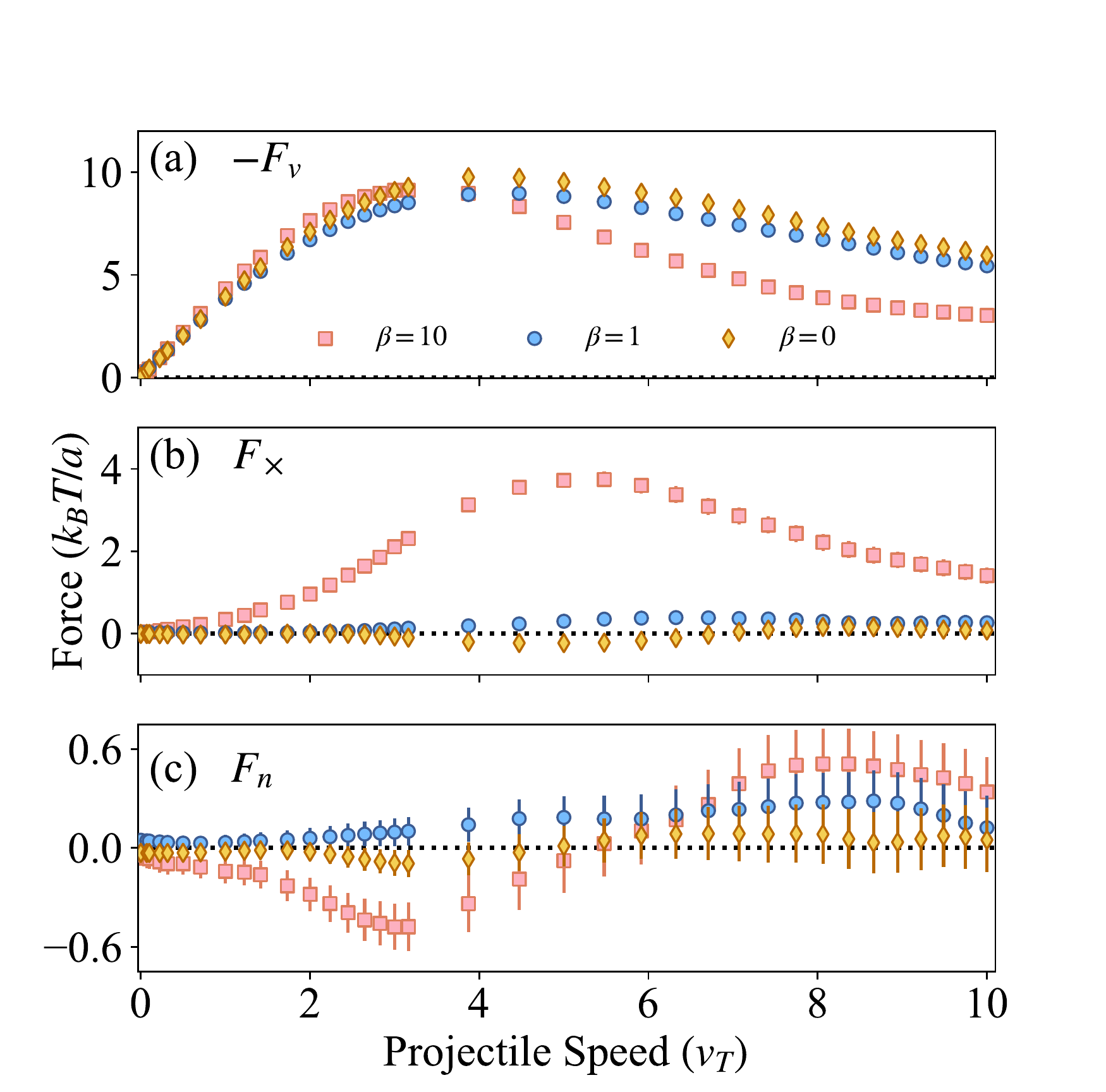}
\caption{Friction force components [$-F_v$ in panel (a), $F_\times$ in (b), and $F_n$ in (c)] for $\Gamma = 10$, $\theta = 22.5^\circ$, and $\beta = 0$ (diamonds), 1 (circles), and 10 (squares).
}
\label{fig:G10_diff_b}
\end{figure}

The transverse ($F_\times$) and Lorentz-directed ($F_n$) components of the friction force are only present when the plasma is strongly magnetized ($\beta  > 1$).
This is demonstrated in Fig.~\ref{fig:G10_diff_b}, where the friction is calculated for a test particle in a plasma with $\Gamma = 10$, $\theta = 22.5^\circ$, and $\beta = 0$, 1, and 10. 
The $F_\times$ component is only non-negligible when $\beta > 1$ [Fig.~\ref{fig:G10_diff_b}(b)], which is similar to predictions in the weakly coupled limit \cite{Lafleur_Baalrud_2019,Jose_Baalrud_2020,bldb_2020}.
However, the $F_n$ component is only non-negligible when both $\Gamma > 1$ and $\beta > 1$ [Fig.~\ref{fig:G10_diff_b}(c)]; its presence appears to arise from the combination of strong coupling and strong magnetization, and not one of these conditions alone.

It is also interesting to notice that the stopping power component in the strongly coupled regime depends on $\beta$ in a qualitatively similar way as at weakly coupling~\cite{Lafleur_Baalrud_2019,Mag_dEdx_Book}. 
In particular, as $\beta$ increases the Bragg peak shifts to lower speeds and the high speed stopping decreases more rapidly with speed [Fig.~\ref{fig:G10_diff_b}(a)]. 
Strong magnetization causes an increase in the stopping power at low speed, but a decrease at high speed.

 \subsection{Influence of angle} 
 
\begin{figure}
\centering
\includegraphics[width=8.5cm]{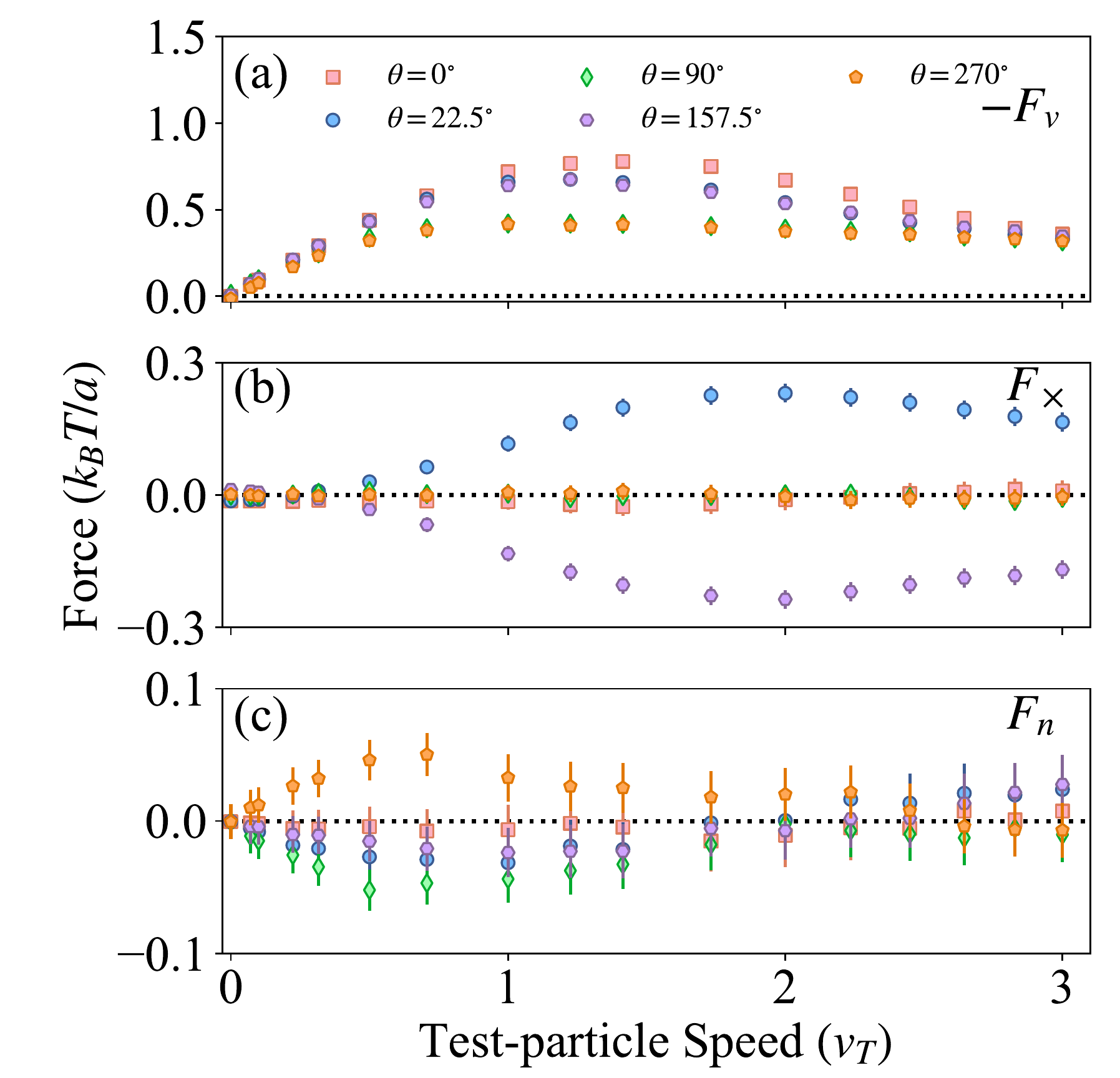}
\caption{Friction force components [$-F_v$ in panel (a), $F_\times$ in (b), and $F_n$ in (c)] at $\Gamma=1$, $\beta = 10$ and five angles: $\theta= 0^\circ$ (squares), $22.5^\circ$ (circles), $90^\circ$ (diamonds), $157.5^\circ$ (hexagons), and $270^\circ$ (pentagons).
}
\label{fig:angles}
\end{figure}

The friction force also depends significantly on the angle between the velocity and magnetic field, $\theta$.
An example is presented in Fig.~\ref{fig:angles}, which shows results for  $\Gamma = 1$, $\beta = 10$, and $\theta = 0^\circ$, $22.5^\circ$, $90^\circ$, $157.5^\circ$, and $270^\circ$.
Some qualitative features are similar to expectations from linear response theory at weak coupling.~\cite{Lafleur_Baalrud_2019,bldb_2020,Mag_dEdx_Book}.
For example, the peak of the stopping power component ($F_v$) shifts to lower speed and decreases in magnitude when the test particle moves perpendicular to the magnetic field ($\theta = 90^\circ$ and $270^\circ$). [Fig.~\ref{fig:angles}(a)]. 
Expected symmetries in the stopping power component are also observed, as the data for $22.5^\circ$ and $157.5^\circ$ give the same values, as do those at $90^\circ$ and $270^\circ$. 
The stopping power component is expected to have a $F_v(\theta) = F_v(\pi + \theta)$ and $F_v(\theta) = F_v( - \theta)$ symmetry. 

Expected symmetry properties are also confirmed in the transverse component ($F_\times$) [Fig.~\ref{fig:angles}(c)]. 
It is zero when the particle moves parallel ($\theta = 0^\circ$), or perpendicular ($\theta = 90^\circ$ and $270^\circ$) to the magnetic field. 
It is also equal in magnitude but opposite in sign when $\theta = 22.5^\circ$ and $\theta = 157.5^\circ$.
The symmetries $F_\times(\theta) = F_\times(\pi + \theta)$, and $F_\times(\theta) = -F_\times(\pi - \theta)$ are predicted by linear response theory~\cite{Lafleur_Baalrud_2019}, and binary collision theory~\cite{Jose_Baalrud_2020}, which is consistent with the MD data. 

The component of the friction force in the direction of the Lorentz force, $F_n$, is observed to have different symmetry properties than the other directions [Fig.~\ref{fig:angles}(c)].
It appears to have maximal values when the test-particle's velocity is perpendicular to the magnetic field ($\theta = 90^\circ$ and $270^\circ$). 
It is also observed that $F_n(22.5^\circ) \approx F_n(157.5^\circ)$. 
Although limited, this data seems to suggest that $F_n$ obeys the symmetry properties $F_n(\theta) = -F_n(\pi + \theta)$ and $F_n(\theta) = F_n(\pi - \theta)$.
This translates to a consistent symmetry as the $\sin \theta$ dependence of the Lorentz force. 
A consequence is that a positive sign of $F_n$ in the first quadrant ($\theta = 0-90^\circ$) will translate to a force that increases the gyrofrequency of particle in all quadrants; i.e., independent of the phase angle $\theta$. 
Conversely, a negative sign of $F_n$ in the first quadrant will translate to a force that decreases the gyrofrequency, independent of the phase angle $\theta$. 

\section{Discussion}
\label{sec:discussion}

\subsection{Potential Wakes}
\label{sec:Wakes}

\begin{figure}
\centering
\includegraphics[width=8cm]{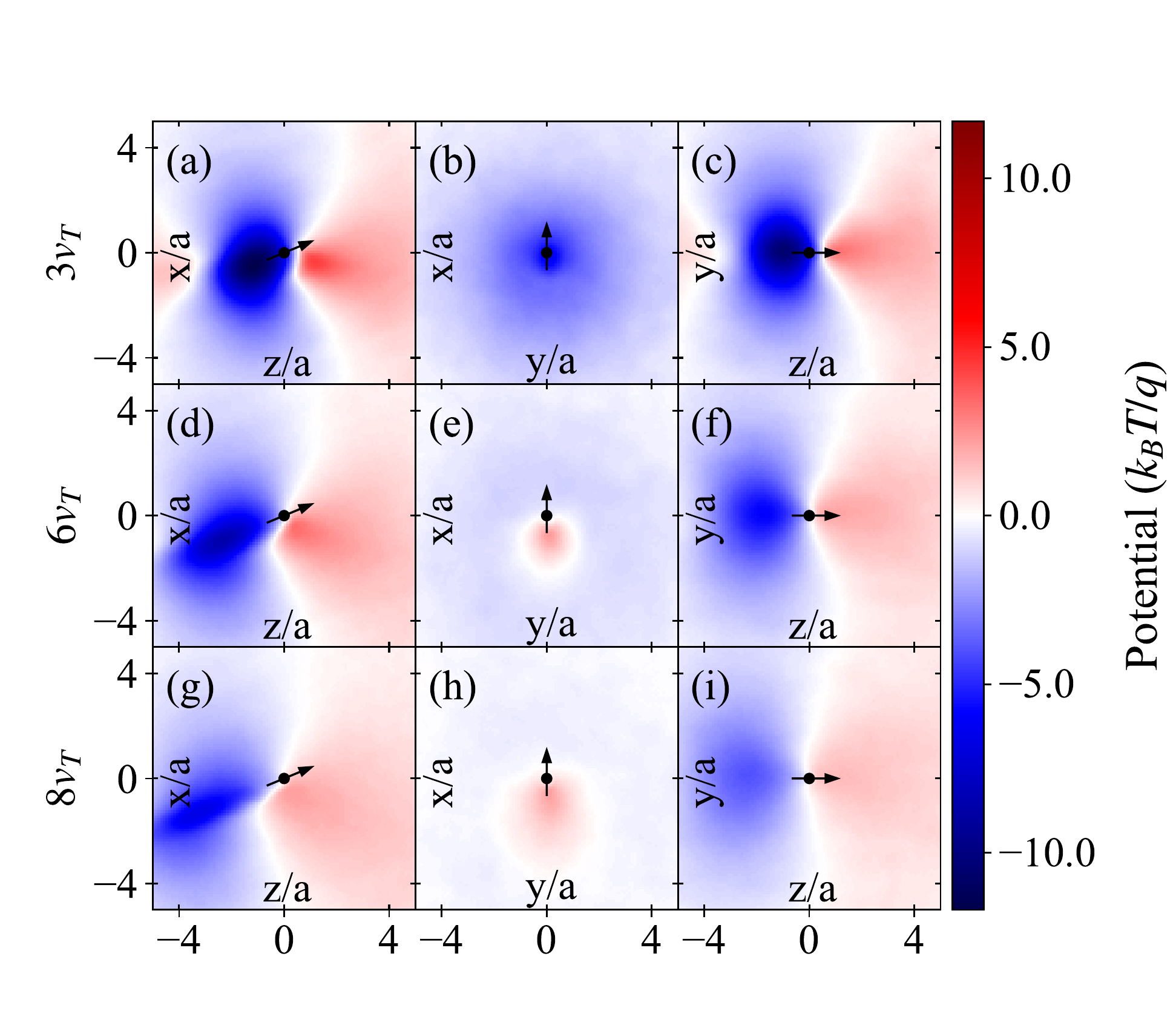}
\caption{Average electrostatic potential distributions in units of $k_BT/q$ about a test-particle traveling in a plasma with $\beta = 10$ ($\vc{B}$ is along the $z-$direction), $\Gamma = 10$, at an angle $\theta=22.5^\circ$.
Each column corresponds to a different plane in a Cartesian coordinate system that contains the test-particle (the $x-z$ plane in the left-most column [panels (a), (d), (g)], the $x-y$ plane in the middle column [panels (b), (e), (h)], and the $y-z$ plane in the right-most column[panels (c), (f), (i)]).
Wakes were calculated for three different speeds (one for each row); $3v_T$ [panels (a), (b), (c)], $6v_T$ [panels (d), (e), (f)], and $8v_T$ [panels (g), (h), (i)].
Arrows show the orientation of the velocity.
}
\label{fig:wakes}
\end{figure}

The friction force on a moving test particle is the electrostatic force exerted by the charge density perturbations induced in its wake~\cite{Ichimaru,Nicholson}.
In an unmagnetized plasma, the wake is symmetric about the velocity of the test charge. 
As a result, the only component of the friction force is aligned antiparallel to the velocity, resulting in the stopping power. 
However, wakes are significantly influenced by strong magnetization, which causes them to rotate toward the direction of the magnetic field~\cite{Shukla_Salimullah_1996,Darian_Miloch_2019,Piel_Greiner_2018,Ware_Wiley_1993,Joost_Ludwig_2014}.
Asymmetries in the wake about the test particle's velocity give rise to the different components of the friction.
Models for the wake potential are usually based on linear response descriptions that do not account for strong coupling. 
The MD simulation results shown in Fig.~\ref{fig:wakes} reveal that wakes persist when the plasma is strongly coupled. 
The symmetry properties of these wakes can be used to visualize what causes each of the three components of the friction force. 

The potential distributions in Fig.~\ref{fig:wakes} were calculated as follows.
A separate set of 30,000 simulations were conducted similarly to those described in Sect.~\ref{sec:sims}.
However, the step at which the test particle maintained fixed momentum was extended to $3.5 \omega_p^{-1}$ rather than $2 \omega_p^{-1}$.
Particle positions were recorded at 2, 2.5, 3, and $3.5 \omega_p^{-1}$ during this step.
For these four timesteps, a potential wake was calculated for each of the 30,000 simulations by first creating a $100 \times 100$ grid of points about the test-particle's position in either the $x-z$, $x-y$, or $y-z$ plane that extended $5a$ away from the test particle (creating a square grid with an edge length of $10a$).
At each grid point, the total Coulomb potential from all the particles within $29a$ of the grid point's location were then calculated.
The 30,000 grids were then averaged yielding one grid per each of the four timesteps.
The four grids for each timestep were then averaged yielding one final grid. 
Finally, the neutralizing background was accounted for by subtracting the potential due to the uniform neutralizing background in each sphere $\Phi_{\textrm{b}} = 3R^2/2\sqrt{\Gamma}$ (where $R = 29a$) from each grid point yielding the final potential distributions in Fig.~\ref{fig:wakes}.

The stopping power component is due to the asymmetry along the test particle's velocity.
This is the only asymmetry present when $\beta = 0$ \cite{Lafleur_Baalrud_2019}.
As the test particle's speed increases, the region responsible for the $F_v$ component (a region of low density behind the test particle) is displaced further from the test particle, decreasing the magnitude of the stopping power.

Because the wake rotates into the direction of $\vc{B}$ in the strongly magnetized regime, there is an asymmetry in the $x-z$ with respect to the test-particle's velocity [Figs.~\ref{fig:wakes}(a), (d), and (g)].
This asymmetry is responsible for the $F_\times$ component, as was previously predicted and observed in the weakly coupled limit \cite{Lafleur_Baalrud_2019,bldb_2020}.
The $F_\times$ component is a maximum at about $6v_T$  when $\Gamma =10$ [Fig.~\ref{fig:force}(g)].
Corresponding to this, the asymmetry is the most obvious and closest to the test particle at this speed [compare Fig.~\ref{fig:wakes}(d) with Figs.~\ref{fig:wakes}(a) and (g)].

Although not as pronounced as the other two, a third asymmetry with respect to the $y-$axis is also present that gives rise to the $F_n$ component [Figs.~\ref{fig:wakes}(b), (e), and (h)].
The wake is predicted to be symmetric with respect to the $y-$direction in the weakly coupled limit, thus no friction force is exerted in the $\hat{\vc{n}}$ direction~\cite{Lafleur_Baalrud_2019}.
Conversely, at the $\Gamma = 10$ and $\beta=10$ conditions of Fig.~\ref{fig:wakes} when the test particle is traveling with a speed of $3v_T$, the test particle is located within the region of negative potential which is slightly shifted in the $y-$direction [Figs.~\ref{fig:wakes}(b)]. 
This results in a net force in the negative $\hat{\vc{n}}$ direction on the test particle.
However, as the test particle's speed is increased, the negative region is displaced by a positive electrostatic potential.
Since the potential perturbation remains shifted in the positive $y-$direction, this leads to a sign change in the $F_n$ component, which is now along the positive $\hat{\vc{n}}$ direction. 
The sign change in $F_n$ can therefore be connected with changes in the wake that occur as the test particle speed changes.

The fact that the $F_n$ component of the friction is not predicted by linear response theory, and is not observed at weak coupling, provides some insight into the mechanisms that cause it.
The linear response based theory does not account for strong interactions near the turning points (distance of closest approach) in particle interactions~\cite{Lafleur_Baalrud_2019}.
Although negligible in the weakly coupled limit, these strong short-range interactions become dominant when $\Gamma > 1$.
This suggests that the component of the friction force in the Lorentz force direction $F_n$ is associated with strong short-range interactions in the presence of a strong magnetic field. 
Such short-range physics is accounted for in the recent generalized collision operator from Ref.~\onlinecite{Jose_Baalrud_2020}.
An extensions of this theory to strong coupling may be able to model the $F_n$ component.

\subsection{Implications for particle dynamics}

The transverse and Lorentz-directed components are expected to influence single particle dynamics.
This can be seen from the equations of motion of a gyrating test particle [Eq.~(\ref{eq:eom})], which are cast here in a spherical coordinate system such that $\vc{B} = B \hat{z}$, so $v_x = v \sin \theta \cos \phi$, $v_y = v \sin \theta \sin \phi$, and $v_z = v \cos \theta$, where $\theta$ is the polar angle, and $\phi$ is the azimuthal angle
\begin{subequations}
\begin{align}
    \frac{d v}{dt} &= - \frac{F_v(v,\theta)}{M} \label{eq:sub:1}\\ 
    \frac{d \theta}{dt} &= \frac{F_\times (v, \theta)}{Mv} \label{eq:sub:2} \\
    \frac{d \phi}{dt} &= - \omega_{c_t} - \frac{F_n(v,\theta)}{Mv\sin \theta}. \label{eq:sub:3}
\end{align}
\end{subequations}
Here, $\omega_{c_t} = QB/cM$ is the gyrofrequency of the test particle. 
Each component affects single particle dynamics as follows.

The $F_v$ component [Eq.~(\ref{eq:sub:1})] acts to slow the test particle's speed.
This is the only component through which the test particle's energy is dissipated.
Both the test particle's velocity parallel ($V_\parallel$) and perpendicular ($V_\perp$) decrease through the stopping power component.
As such, this component acts to always decrease the test particle's gyroradius. 

When $\beta > 1$, the $F_\times$ component couples the test particle speed $v$ and polar angle $\theta$ [Eq.~(\ref{eq:sub:2})].
This component does not dissipate the test particle's energy, but acts to shift the test particle's momentum between the directions parallel and perpendicular to the magnetic field.
When $\Gamma < 1$, theory predicts $F_\times$ changes sign between sufficiently fast and slow particles \cite{Lafleur_Baalrud_2019,Jose_Baalrud_2020}.
As such, the $F_\times$ acts to decrease $V_\parallel$ and increase $V_\perp$ for sufficiently fast test particles, then decrease $V_\perp$ and increase $V_\parallel$ for sufficiently slow test particles.
However when $\Gamma > 1$, no such sign change is observed (Fig.~\ref{fig:force}).
As a result, $F_\times$ in strongly coupled plasmas only acts to increase $V_\perp$ and decrease $V_\parallel$, which acts to increase the test particle's gyroradius.

The equations of motion also show that the rate of gyration changes between strongly magnetized plasmas that are either weakly or strongly coupled [Eq.~(\ref{eq:sub:3})]. 
When $\Gamma < 1$ and $F_n \approx 0$, the test particle gyrates at a constant rate; its gyrofrequency $\omega_{c_t}$.
However when $\Gamma > 1$, $F_n$ couples the azimuthal direction to the test particle speed [Eq.~(\ref{eq:sub:3})].
As mentioned, $F_n$ appears to have the same symmetry with $\theta$ as $\sin \theta$, so $F_n / \sin \theta$ likely has the same sign as $\theta$ varies.
Thus, $F_n>0$ increases the rate of gyration, while $F_n<0$ decreases the rate of gyration. 

The effects of strong coupling and strong magnetization on particle dynamics may have implications for experiments that rely on particle confinement, such as those on antimatter traps that necessitate long particle confinement times. 
With an increase in gyroradius via the $F_\times$ component, particles may exit a target confinement volume at a faster rate in strongly coupled and strongly magnetized plasmas than previously expected.
The coupling of particle momentum parallel and perpendicular to the magnetic field via $F_\times$ may also affect macroscopic transport in strongly coupled and strongly magnetized plasmas.
The friction is linked to macroscopic transport, as has been previously examined when $\beta = 0$ \cite{bbd_2019,Dufty1,Dufty2}.
Likewise, the friction force is related to electrical conductivity \cite{Braginskii}.
The $F_\times$ component couples the parallel and perpendicular collisions through collisions in a way that is not present at weak magnetization~\cite{Braginskii}.
This could provide a possible mechanism for deviations between simulation calculations of macroscopic transport quantities from the trends predicted by conventional theory \cite{Baalrud_Daligault_MagPhases,Ott_Bonitz_PRL,Okuda_Dawson_PRL}.
Moreover, $F_n$ couples the test particle's speed with the rate of gyration, providing another mechanism that may influence transport.

\section{Conclusion}
\label{sec:conclusion}

These results show that the transverse friction force that was previously observed to arise due to strong magnetization in weakly coupled plasmas becomes larger in both absolute and relative terms in the regime of strong Coulomb coupling.
Furthermore, the combination of strong magnetization and strong coupling is found to lead to a new effect where the friction force has a component in the direction of the Lorentz force. 
Although this is small compared to the other two components, it is a qualitatively new contribution that can influence the gyrofrequency of a test particle as it traverses a plasma. 

These new behaviors associated with strong magnetization inform the development of kinetic theory.
For example, a component of the friction in the Lorentz force direction is not predicted by the previous linear response theory for weakly coupled plasmas \cite{Lafleur_Baalrud_2019}. 
Although this is consistent with the MD simulations in the weakly coupled regime \cite{bldb_2020}, extensions of linear response theory that have been proposed to treat strong coupling effects using static local field corrections~\cite{Ichimaru, Zwick_Review} would still possess the same symmetry property that leads to the prediction that $F_n=0$, as described in Ref.~\onlinecite{Lafleur_Baalrud_2019}. 
This suggests that strong short-range interactions, which are neglected in linear response theory, are responsible for $F_n$. 
The recent collisional kinetic theory from Ref.~\onlinecite{Jose_Baalrud_2020}, which was able to capture the transverse friction force at weak coupling, presents a possible avenue to treat moderate-to-strong coupling along with strong magnetization. 

These results also imply that novel physical effects associated with strong magnetization should be expected at the level of macroscopic plasma transport. 
For instance, the electrical resistivity coefficient is directly related to the friction force between ions and electrons, so the transverse and Lorentz-directed friction forces will influence the tensor resistivity coefficients in a way that is qualitatively different than in weakly magnetized plasmas (where these components do not exist). 
Other examples of macroscopic transport that may be affected are self-diffusion and thermal relaxation, where links between the friction and these coefficients have been shown when $\beta = 0$ \cite{bbd_2019,Dufty1,Dufty2}.
Such effects should be expected to arise in strongly magnetized plasmas found in experiments and natural systems,
such as antimatter traps \cite{Surko_Fajan,antimatter_2015,antimatter_2004}, nonneutral plasmas, \cite{nonneutral_PRL_1977,nonneutral_PRL_1980,nonneutral_PRL_1980,nonneutral_PhysFluids_1980}, and ultracold neutral plasmas \cite{Ultracold_2,Killian_UCNP}, and neutron star atmospheres \cite{neutron_star_1}.
It suggests that these systems access a regime for which there is little theoretical basis to understand transport. 
These are interesting platforms for exploring fundamental new regimes of plasma physics.  

\section{Data Availability} 

The data that support the findings of this study are available in the supplementary materials document.

\begin{acknowledgements}

The authors thank Dr.~Jerome Daligault for supplying the MD code used in this work, and for helpful discussions. 
We also thank Dr.~Trevor Lafleur for helpful discussions. 

This material is based upon work supported by the U.S. Department of Energy, Office of Science, Office of Fusion Energy Sciences under Award Number DE-SC0016159, the U.S. Department of Energy, National Nuclear Security Administration, under Award Number DE-NA0003868, and by the National Science Foundation under Grant No.~PHY-1453736. 
It used the Extreme Science and Engineering Discovery Environment (XSEDE), which is supported by NSF Grant No. ACI-1053575, under Project Award No. PHYS-150018. 
\end{acknowledgements}

\bibliography{refs.bib}

\end{document}


\preprint{AIP/123-QED}
\begin{table}
\begin{adjustwidth}{-2.5cm}{}
\begin{subtable}{0.81 \textwidth} \centering
{\begin{tabular}{ccccccc}
\hline
\hline
$V$ & $-F_v$ $(\times 10^{-2})$ & $\sigma_{F_v}$ $(\times 10^{-2})$ & $F_\times$ $(\times 10^{-2})$ &$\sigma_{F_\times}$ $(\times 10^{-2})$ & $F_n$ $(\times 10^{-2})$& $\sigma_{F_n}$ $(\times 10^{-2})$ \\
\hline
    0.0000 &    -0.0241 &     0.0861 &    -0.1031 &     0.0882 &     0.0364 &    -0.0854 \\ 
    0.0707 &     0.3647 &     0.0851 &    -0.2237 &     0.0862 &     0.0333 &    -0.0860 \\ 
    0.1000 &     0.5521 &     0.0853 &    -0.1961 &     0.0864 &     0.0223 &    -0.0873 \\ 
    0.2236 &     0.9546 &     0.0847 &    -0.0664 &     0.0862 &    -0.0390 &    -0.0873 \\ 
    0.3162 &     1.3178 &     0.0846 &     0.0430 &     0.0860 &     0.0819 &    -0.0894 \\ 
    0.5000 &     1.9285 &     0.0865 &     0.1461 &     0.0863 &    -0.0352 &    -0.0884 \\ 
    0.7071 &     2.0196 &     0.0818 &     0.2648 &     0.0849 &     0.1382 &    -0.0889 \\ 
    1.0000 &     2.2090 &     0.0800 &     0.2318 &     0.0852 &     0.0831 &    -0.0895 \\ 
    1.2247 &     2.0201 &     0.0773 &     0.3797 &     0.0847 &    -0.0920 &    -0.0882 \\ 
    1.4142 &     1.8824 &     0.0759 &     0.5116 &     0.0852 &     0.0180 &    -0.0865 \\ 
    1.7321 &     1.6906 &     0.0734 &     0.2824 &     0.0833 &    -0.0768 &    -0.0870 \\ 
    2.0000 &     1.5565 &     0.0716 &     0.3640 &     0.0830 &    -0.1335 &    -0.0866 \\ 
    2.2361 &     1.2701 &     0.0651 &     0.3373 &     0.0782 &     0.0626 &    -0.0847 \\ 
    2.4495 &     1.1477 &     0.0628 &     0.2415 &     0.0774 &     0.1000 &    -0.0844 \\ 
    2.6458 &     1.0390 &     0.0610 &     0.2223 &     0.0778 &    -0.0318 &    -0.0828 \\ 
    2.8284 &     0.9213 &     0.0593 &     0.2204 &     0.0742 &    -0.0001 &    -0.0804 \\ 
    3.0000 &     0.8205 &     0.0554 &     0.2545 &     0.0710 &     0.1252 &    -0.0787 \\
\hline
\hline
 \end{tabular}}
\caption{Force data for $\Gamma = 0.1$, $\beta = 10$, $\theta = 22.5^\circ$.}
\end{subtable}%
\begin{subtable}{0.5\textwidth} \centering
{\begin{tabular}{ccccccc}
\hline
\hline
$V$ & $-F_v$ & $\sigma_{F_v}$ & $F_\times$ & $\sigma_{F_\times}$ & $F_n$ & $\sigma_{F_n}$ \\
\hline
    0.0000 &    -0.0014 &     0.0051 &    -0.0139 &     0.0051 &    -0.0005 &    -0.0051 \\ 
    0.0707 &     0.0666 &     0.0051 &    -0.0139 &     0.0051 &    -0.0014 &    -0.0052 \\ 
    0.1000 &     0.0940 &     0.0051 &    -0.0138 &     0.0051 &    -0.0019 &    -0.0052 \\ 
    0.2236 &     0.2087 &     0.0053 &    -0.0144 &     0.0053 &    -0.0057 &    -0.0054 \\ 
    0.3162 &     0.2911 &     0.0054 &    -0.0120 &     0.0054 &    -0.0059 &    -0.0055 \\ 
    0.5000 &     0.4409 &     0.0058 &    -0.0159 &     0.0059 &    -0.0042 &    -0.0059 \\ 
    0.7071 &     0.5805 &     0.0063 &    -0.0136 &     0.0064 &    -0.0076 &    -0.0065 \\ 
    1.0000 &     0.7196 &     0.0072 &    -0.0150 &     0.0073 &    -0.0066 &    -0.0073 \\ 
    1.2247 &     0.7673 &     0.0078 &    -0.0212 &     0.0077 &    -0.0017 &    -0.0079 \\ 
    1.4142 &     0.7786 &     0.0083 &    -0.0266 &     0.0082 &    -0.0045 &    -0.0082 \\ 
    1.7321 &     0.7508 &     0.0088 &    -0.0203 &     0.0089 &    -0.0152 &    -0.0089 \\ 
    2.0000 &     0.6714 &     0.0090 &    -0.0106 &     0.0093 &    -0.0108 &    -0.0094 \\ 
    2.2361 &     0.5891 &     0.0091 &    -0.0048 &     0.0093 &    -0.0050 &    -0.0095 \\ 
    2.4495 &     0.5154 &     0.0089 &     0.0035 &     0.0094 &    -0.0056 &    -0.0095 \\ 
    2.6458 &     0.4502 &     0.0086 &     0.0066 &     0.0093 &     0.0075 &    -0.0093 \\ 
    2.8284 &     0.3941 &     0.0082 &     0.0133 &     0.0093 &     0.0008 &    -0.0092 \\ 
    3.0000 &     0.3591 &     0.0079 &     0.0095 &     0.0092 &     0.0076 &    -0.0092 \\
\hline
\hline
\label{tab:G1_b10_t0}
 \end{tabular}}
 \caption{Force data for $\Gamma = 1$, $\beta = 10$, $\theta = 0^\circ$.}
 \end{subtable}%
\end{adjustwidth}

\begin{subtable}{0.52\textwidth} \centering
{\begin{tabular}{ccccccc}
\hline
\hline
$V$ & $-F_v$ & $\sigma_{F_v}$ & $F_\times$ & $\sigma_{F_\times}$ & $F_n$ & $\sigma_{F_n}$ \\
\hline
    0.0000 &     0.0040 &     0.0051 &    -0.0134 &     0.0051 &    -0.0005 &    -0.0051 \\ 
    0.0707 &     0.0713 &     0.0051 &    -0.0109 &     0.0051 &    -0.0056 &    -0.0052 \\ 
    0.1000 &     0.0971 &     0.0051 &    -0.0096 &     0.0051 &    -0.0081 &    -0.0052 \\ 
    0.2236 &     0.2099 &     0.0053 &    -0.0030 &     0.0052 &    -0.0183 &    -0.0054 \\ 
    0.3162 &     0.2903 &     0.0054 &     0.0087 &     0.0054 &    -0.0207 &    -0.0055 \\ 
    0.5000 &     0.4316 &     0.0058 &     0.0301 &     0.0058 &    -0.0271 &    -0.0059 \\ 
    0.7071 &     0.5615 &     0.0064 &     0.0635 &     0.0064 &    -0.0290 &    -0.0064 \\ 
    1.0000 &     0.6596 &     0.0072 &     0.1162 &     0.0071 &    -0.0315 &    -0.0072 \\ 
    1.2247 &     0.6752 &     0.0076 &     0.1643 &     0.0076 &    -0.0188 &    -0.0076 \\ 
    1.4142 &     0.6572 &     0.0078 &     0.1981 &     0.0078 &    -0.0215 &    -0.0079 \\ 
    1.7321 &     0.6139 &     0.0082 &     0.2259 &     0.0082 &    -0.0014 &    -0.0084 \\ 
    2.0000 &     0.5414 &     0.0082 &     0.2308 &     0.0083 &     0.0004 &    -0.0085 \\ 
    2.2361 &     0.4807 &     0.0082 &     0.2216 &     0.0084 &     0.0164 &    -0.0086 \\ 
    2.4495 &     0.4276 &     0.0079 &     0.2098 &     0.0084 &     0.0136 &    -0.0087 \\ 
    2.6458 &     0.3921 &     0.0079 &     0.1933 &     0.0084 &     0.0210 &    -0.0087 \\ 
    2.8284 &     0.3558 &     0.0076 &     0.1779 &     0.0083 &     0.0197 &    -0.0087 \\ 
    3.0000 &     0.3294 &     0.0075 &     0.1653 &     0.0083 &     0.0238 &    -0.0086 \\ 
\hline
\hline
\label{tab:G1_b10_t22p5}
 \end{tabular}}
 \caption{Force data for $\Gamma = 1$, $\beta = 10$, $\theta = 22.5^\circ$.}
 \end{subtable}%
\begin{subtable}{0.5\textwidth} \centering
{\begin{tabular}{ccccccc}
\hline
\hline
$V$ & $-F_v$ & $\sigma_{F_v}$ & $F_\times$ & $\sigma_{F_\times}$ & $F_n$ & $\sigma_{F_n}$ \\
\hline
    0.0000 &     0.0139 &     0.0051 &    -0.0014 &     0.0051 &    -0.0005 &    -0.0051 \\ 
    0.0707 &     0.0740 &     0.0051 &    -0.0021 &     0.0052 &    -0.0113 &    -0.0052 \\ 
    0.1000 &     0.0983 &     0.0051 &    -0.0014 &     0.0052 &    -0.0150 &    -0.0053 \\ 
    0.2236 &     0.1922 &     0.0053 &     0.0011 &     0.0054 &    -0.0257 &    -0.0054 \\ 
    0.3162 &     0.2478 &     0.0054 &     0.0052 &     0.0056 &    -0.0346 &    -0.0056 \\ 
    0.5000 &     0.3382 &     0.0057 &     0.0092 &     0.0061 &    -0.0521 &    -0.0060 \\ 
    0.7071 &     0.3933 &     0.0059 &     0.0022 &     0.0065 &    -0.0469 &    -0.0064 \\ 
    1.0000 &     0.4196 &     0.0061 &     0.0030 &     0.0070 &    -0.0438 &    -0.0069 \\ 
    1.2247 &     0.4199 &     0.0062 &    -0.0070 &     0.0073 &    -0.0374 &    -0.0071 \\ 
    1.4142 &     0.4178 &     0.0062 &    -0.0025 &     0.0074 &    -0.0327 &    -0.0073 \\ 
    1.7321 &     0.4050 &     0.0062 &    -0.0022 &     0.0077 &    -0.0178 &    -0.0075 \\ 
    2.0000 &     0.3901 &     0.0063 &     0.0019 &     0.0078 &    -0.0023 &    -0.0077 \\ 
    2.2361 &     0.3780 &     0.0064 &     0.0032 &     0.0080 &    -0.0072 &    -0.0079 \\ 
    2.4495 &     0.3690 &     0.0065 &    -0.0048 &     0.0081 &    -0.0098 &    -0.0080 \\ 
    2.6458 &     0.3449 &     0.0063 &    -0.0110 &     0.0082 &    -0.0127 &    -0.0080 \\ 
    2.8284 &     0.3336 &     0.0062 &    -0.0135 &     0.0082 &    -0.0054 &    -0.0081 \\ 
    3.0000 &     0.3166 &     0.0059 &    -0.0065 &     0.0081 &    -0.0102 &    -0.0080 \\
\hline
\hline
 \end{tabular}}
\caption{Force data for $\Gamma = 1$, $\beta = 10$, $\theta = 90^\circ$.}
\end{subtable}%

\begin{subtable}{0.52\textwidth} \centering
{\begin{tabular}{ccccccc}
\hline
\hline
$V$ & $-F_v$ & $\sigma_{F_v}$ & $F_\times$ & $\sigma_{F_\times}$ & $F_n$ & $\sigma_{F_n}$ \\
\hline
    0.0000 &     0.0066 &     0.0051 &     0.0123 &     0.0051 &    -0.0005 &    -0.0051 \\ 
    0.0707 &     0.0734 &     0.0052 &     0.0090 &     0.0051 &    -0.0041 &    -0.0052 \\ 
    0.1000 &     0.1021 &     0.0052 &     0.0065 &     0.0051 &    -0.0043 &    -0.0052 \\ 
    0.2236 &     0.2131 &     0.0053 &    -0.0007 &     0.0053 &    -0.0102 &    -0.0054 \\ 
    0.3162 &     0.2922 &     0.0055 &    -0.0089 &     0.0055 &    -0.0108 &    -0.0055 \\ 
    0.5000 &     0.4316 &     0.0059 &    -0.0329 &     0.0058 &    -0.0153 &    -0.0060 \\ 
    0.7071 &     0.5457 &     0.0064 &    -0.0671 &     0.0063 &    -0.0208 &    -0.0065 \\ 
    1.0000 &     0.6391 &     0.0071 &    -0.1329 &     0.0070 &    -0.0238 &    -0.0072 \\ 
    1.2247 &     0.6744 &     0.0077 &    -0.1747 &     0.0075 &    -0.0228 &    -0.0077 \\ 
    1.4142 &     0.6411 &     0.0078 &    -0.2037 &     0.0077 &    -0.0228 &    -0.0080 \\ 
    1.7321 &     0.6007 &     0.0081 &    -0.2281 &     0.0081 &    -0.0055 &    -0.0083 \\ 
    2.0000 &     0.5361 &     0.0081 &    -0.2365 &     0.0082 &    -0.0072 &    -0.0085 \\ 
    2.2361 &     0.4851 &     0.0081 &    -0.2189 &     0.0084 &     0.0017 &    -0.0086 \\ 
    2.4495 &     0.4367 &     0.0079 &    -0.2032 &     0.0083 &     0.0021 &    -0.0086 \\ 
    2.6458 &     0.3983 &     0.0077 &    -0.1879 &     0.0083 &     0.0134 &    -0.0087 \\ 
    2.8284 &     0.3783 &     0.0077 &    -0.1826 &     0.0084 &     0.0216 &    -0.0087 \\ 
    3.0000 &     0.3474 &     0.0076 &    -0.1691 &     0.0083 &     0.0276 &    -0.0087 \\
\hline
\hline
 \end{tabular}}
\caption{Force data for $\Gamma = 1$, $\beta = 10$, $\theta = 157.5^\circ$.}
\end{subtable}%
\begin{subtable}{0.5\textwidth} \centering
{\begin{tabular}{ccccccc}
\hline
\hline
$V$ & $-F_v$ & $\sigma_{F_v}$ & $F_\times$ & $\sigma_{F_\times}$ & $F_n$ & $\sigma_{F_n}$ \\
\hline
    0.0000 &    -0.0139 &     0.0051 &     0.0014 &     0.0051 &    -0.0005 &    -0.0051 \\ 
    0.0707 &     0.0501 &     0.0052 &    -0.0002 &     0.0051 &     0.0102 &    -0.0052 \\ 
    0.1000 &     0.0755 &     0.0052 &    -0.0023 &     0.0052 &     0.0121 &    -0.0052 \\ 
    0.2236 &     0.1690 &     0.0053 &     0.0024 &     0.0054 &     0.0265 &    -0.0054 \\ 
    0.3162 &     0.2333 &     0.0054 &    -0.0024 &     0.0055 &     0.0321 &    -0.0055 \\ 
    0.5000 &     0.3210 &     0.0056 &     0.0011 &     0.0060 &     0.0461 &    -0.0060 \\ 
    0.7071 &     0.3811 &     0.0059 &    -0.0005 &     0.0065 &     0.0502 &    -0.0064 \\ 
    1.0000 &     0.4162 &     0.0062 &     0.0060 &     0.0069 &     0.0327 &    -0.0070 \\ 
    1.2247 &     0.4087 &     0.0061 &     0.0030 &     0.0072 &     0.0262 &    -0.0072 \\ 
    1.4142 &     0.4138 &     0.0063 &     0.0091 &     0.0074 &     0.0251 &    -0.0073 \\ 
    1.7321 &     0.3971 &     0.0063 &     0.0028 &     0.0078 &     0.0181 &    -0.0076 \\ 
    2.0000 &     0.3750 &     0.0062 &    -0.0037 &     0.0079 &     0.0202 &    -0.0077 \\ 
    2.2361 &     0.3626 &     0.0061 &    -0.0114 &     0.0079 &     0.0218 &    -0.0078 \\ 
    2.4495 &     0.3569 &     0.0061 &    -0.0076 &     0.0081 &     0.0079 &    -0.0079 \\ 
    2.6458 &     0.3397 &     0.0059 &    -0.0085 &     0.0081 &    -0.0040 &    -0.0080 \\ 
    2.8284 &     0.3301 &     0.0060 &    -0.0075 &     0.0081 &    -0.0065 &    -0.0079 \\ 
    3.0000 &     0.3183 &     0.0059 &    -0.0042 &     0.0080 &    -0.0069 &    -0.0080 \\ 

\hline
\hline
 \end{tabular}}
\caption{Force data for $\Gamma = 1$, $\beta = 10$, $\theta = 270^\circ$.}
\label{tab:G1_b10_t270}
\end{subtable}%
\end{table}

\begin{table}
\begin{subtable}{0.52\textwidth} \centering
{\begin{tabular}{ccccccc}
\hline
\hline
$V$ & $-F_v$ & $\sigma_{F_v}$ & $F_\times$ & $\sigma_{F_\times}$ & $F_n$ & $\sigma_{F_n}$ \\
\hline
    0.0000 &     0.0334 &     0.0142 &    -0.0040 &     0.0141 &    -0.0301 &    -0.0141 \\ 
    0.0707 &     0.3215 &     0.0142 &    -0.0067 &     0.0141 &    -0.0304 &    -0.0141 \\ 
    0.1000 &     0.4407 &     0.0142 &    -0.0077 &     0.0141 &    -0.0304 &    -0.0141 \\ 
    0.2236 &     0.9416 &     0.0143 &    -0.0119 &     0.0142 &    -0.0306 &    -0.0142 \\ 
    0.3162 &     1.3144 &     0.0144 &    -0.0147 &     0.0143 &    -0.0305 &    -0.0143 \\ 
    0.5000 &     2.0448 &     0.0147 &    -0.0189 &     0.0146 &    -0.0296 &    -0.0147 \\ 
    0.7071 &     2.8488 &     0.0151 &    -0.0211 &     0.0151 &    -0.0272 &    -0.0152 \\ 
    1.0000 &     3.9406 &     0.0161 &    -0.0195 &     0.0160 &    -0.0216 &    -0.0162 \\ 
    1.2247 &     4.7345 &     0.0171 &    -0.0150 &     0.0169 &    -0.0168 &    -0.0171 \\ 
    1.4142 &     5.3697 &     0.0181 &    -0.0095 &     0.0177 &    -0.0138 &    -0.0180 \\ 
    1.7321 &     6.3552 &     0.0201 &     0.0007 &     0.0195 &    -0.0145 &    -0.0198 \\ 
    2.0000 &     7.0987 &     0.0221 &     0.0067 &     0.0213 &    -0.0233 &    -0.0216 \\ 
    2.2361 &     7.6825 &     0.0240 &     0.0061 &     0.0230 &    -0.0373 &    -0.0233 \\ 
    2.4495 &     8.1498 &     0.0259 &    -0.0024 &     0.0247 &    -0.0535 &    -0.0250 \\ 
    2.6458 &     8.5273 &     0.0277 &    -0.0185 &     0.0264 &    -0.0689 &    -0.0267 \\ 
    2.8284 &     8.8334 &     0.0294 &    -0.0411 &     0.0280 &    -0.0819 &    -0.0283 \\ 
    3.0000 &     9.0814 &     0.0311 &    -0.0677 &     0.0296 &    -0.0906 &    -0.0299 \\ 
    3.1623 &     9.2804 &     0.0326 &    -0.0961 &     0.0311 &    -0.0944 &    -0.0314 \\ 
    3.8730 &     9.7545 &     0.0391 &    -0.2008 &     0.0383 &    -0.0660 &    -0.0382 \\ 
    4.4721 &     9.7246 &     0.0431 &    -0.2254 &     0.0440 &    -0.0272 &    -0.0435 \\ 
    5.0000 &     9.5223 &     0.0462 &    -0.2277 &     0.0487 &     0.0118 &    -0.0478 \\ 
    5.4772 &     9.2656 &     0.0492 &    -0.2170 &     0.0529 &     0.0460 &    -0.0517 \\ 
    5.9161 &     8.9980 &     0.0519 &    -0.1707 &     0.0566 &     0.0757 &    -0.0556 \\ 
    6.3246 &     8.7485 &     0.0543 &    -0.1064 &     0.0598 &     0.0850 &    -0.0592 \\ 
    6.7082 &     8.4809 &     0.0560 &    -0.0276 &     0.0627 &     0.0869 &    -0.0624 \\ 
    7.0711 &     8.2063 &     0.0574 &     0.0461 &     0.0648 &     0.0873 &    -0.0649 \\ 
    7.4162 &     7.9137 &     0.0583 &     0.0978 &     0.0667 &     0.0841 &    -0.0671 \\ 
    7.7460 &     7.6071 &     0.0583 &     0.1407 &     0.0682 &     0.0869 &    -0.0688 \\ 
    8.0623 &     7.3219 &     0.0582 &     0.1632 &     0.0693 &     0.0833 &    -0.0702 \\ 
    8.3666 &     7.0748 &     0.0577 &     0.1679 &     0.0703 &     0.0544 &    -0.0711 \\ 
    8.6603 &     6.8612 &     0.0578 &     0.1556 &     0.0713 &     0.0323 &    -0.0716 \\ 
    8.9443 &     6.6725 &     0.0588 &     0.1399 &     0.0721 &     0.0353 &    -0.0720 \\ 
    9.2195 &     6.5007 &     0.0600 &     0.1234 &     0.0730 &     0.0543 &    -0.0728 \\ 
    9.4868 &     6.3396 &     0.0607 &     0.0995 &     0.0738 &     0.0736 &    -0.0738 \\ 
    9.7468 &     6.1671 &     0.0605 &     0.0888 &     0.0748 &     0.0676 &    -0.0750 \\ 
   10.0000 &     5.9437 &     0.0591 &     0.0791 &     0.0757 &     0.0489 &    -0.0759 \\
\hline
\hline
 \end{tabular}}
\caption{Force data for $\Gamma = 10$, $\beta = 0$, $\theta = 22.5^\circ$.}
\label{tab:G10_b0_t22p5}
\end{subtable}%
\begin{subtable}{0.5\textwidth} \centering
{\begin{tabular}{ccccccc}
\hline
\hline
$V$ & $-F_v$ & $\sigma_{F_v}$ & $F_\times$ & $\sigma_{F_\times}$ & $F_n$ & $\sigma_{F_n}$ \\
\hline
    0.0000 &     0.0766 &     0.0154 &     0.0314 &     0.0160 &     0.0474 &    -0.0161 \\ 
    0.0707 &     0.3565 &     0.0154 &     0.0300 &     0.0160 &     0.0437 &    -0.0161 \\ 
    0.1000 &     0.4723 &     0.0154 &     0.0295 &     0.0160 &     0.0422 &    -0.0162 \\ 
    0.2236 &     0.9590 &     0.0155 &     0.0270 &     0.0161 &     0.0369 &    -0.0163 \\ 
    0.3162 &     1.3209 &     0.0156 &     0.0251 &     0.0162 &     0.0337 &    -0.0164 \\ 
    0.5000 &     2.0280 &     0.0159 &     0.0215 &     0.0166 &     0.0299 &    -0.0167 \\ 
    0.7071 &     2.8016 &     0.0164 &     0.0189 &     0.0171 &     0.0290 &    -0.0173 \\ 
    1.0000 &     3.8413 &     0.0175 &     0.0180 &     0.0182 &     0.0329 &    -0.0182 \\ 
    1.2247 &     4.5853 &     0.0186 &     0.0193 &     0.0192 &     0.0377 &    -0.0192 \\ 
    1.4142 &     5.1702 &     0.0196 &     0.0214 &     0.0203 &     0.0419 &    -0.0202 \\ 
    1.7321 &     6.0566 &     0.0216 &     0.0268 &     0.0221 &     0.0499 &    -0.0221 \\ 
    2.0000 &     6.7069 &     0.0236 &     0.0359 &     0.0239 &     0.0589 &    -0.0240 \\ 
    2.2361 &     7.2067 &     0.0255 &     0.0476 &     0.0257 &     0.0681 &    -0.0257 \\ 
    2.4495 &     7.6002 &     0.0273 &     0.0608 &     0.0274 &     0.0768 &    -0.0274 \\ 
    2.6458 &     7.9130 &     0.0290 &     0.0766 &     0.0291 &     0.0842 &    -0.0291 \\ 
    2.8284 &     8.1617 &     0.0307 &     0.0951 &     0.0308 &     0.0910 &    -0.0307 \\ 
    3.0000 &     8.3586 &     0.0323 &     0.1152 &     0.0323 &     0.0960 &    -0.0323 \\ 
    3.1623 &     8.5130 &     0.0338 &     0.1340 &     0.0338 &     0.1016 &    -0.0338 \\ 
    3.8730 &     8.9073 &     0.0403 &     0.1920 &     0.0403 &     0.1411 &    -0.0403 \\ 
    4.4721 &     8.9670 &     0.0458 &     0.2385 &     0.0456 &     0.1780 &    -0.0456 \\ 
    5.0000 &     8.8162 &     0.0497 &     0.3025 &     0.0500 &     0.1855 &    -0.0506 \\ 
    5.4772 &     8.5603 &     0.0521 &     0.3535 &     0.0534 &     0.1763 &    -0.0544 \\ 
    5.9161 &     8.2734 &     0.0541 &     0.3782 &     0.0562 &     0.1765 &    -0.0575 \\ 
    6.3246 &     7.9780 &     0.0560 &     0.3913 &     0.0586 &     0.2005 &    -0.0602 \\ 
    6.7082 &     7.7045 &     0.0578 &     0.3852 &     0.0606 &     0.2261 &    -0.0625 \\ 
    7.0711 &     7.4281 &     0.0589 &     0.3688 &     0.0624 &     0.2340 &    -0.0648 \\ 
    7.4162 &     7.1717 &     0.0598 &     0.3580 &     0.0639 &     0.2506 &    -0.0667 \\ 
    7.7460 &     6.9300 &     0.0603 &     0.3338 &     0.0655 &     0.2722 &    -0.0682 \\ 
    8.0623 &     6.7183 &     0.0607 &     0.2999 &     0.0667 &     0.2770 &    -0.0693 \\ 
    8.3666 &     6.5139 &     0.0608 &     0.2633 &     0.0678 &     0.2788 &    -0.0705 \\ 
    8.6603 &     6.2983 &     0.0611 &     0.2470 &     0.0686 &     0.2851 &    -0.0715 \\ 
    8.9443 &     6.0800 &     0.0612 &     0.2484 &     0.0689 &     0.2720 &    -0.0723 \\ 
    9.2195 &     5.8953 &     0.0618 &     0.2570 &     0.0690 &     0.2371 &    -0.0731 \\ 
    9.4868 &     5.7326 &     0.0627 &     0.2708 &     0.0693 &     0.1983 &    -0.0736 \\ 
    9.7468 &     5.5837 &     0.0628 &     0.2727 &     0.0696 &     0.1516 &    -0.0744 \\ 
   10.0000 &     5.4377 &     0.0623 &     0.2664 &     0.0700 &     0.1226 &    -0.0752 \\
\hline
\hline
 \end{tabular}}
\caption{Force data for $\Gamma = 10$, $\beta = 1$, $\theta = 22.5^\circ$.}
\label{tab:G10_b1_t22p5}
\end{subtable}%
\end{table}

\begin{table}
\begin{subtable}{0.52\textwidth} \centering
{\begin{tabular}{ccccccc}
\hline
\hline
$V$ & $-F_v$ & $\sigma_{F_v}$ & $F_\times$ & $\sigma_{F_\times}$ & $F_n$ & $\sigma_{F_n}$ \\
\hline
    0.0000 &    -0.0350 &     0.0248 &     0.0193 &     0.0246 &    -0.0426 &    -0.0249 \\ 
    0.0707 &     0.2806 &     0.0248 &     0.0355 &     0.0245 &    -0.0572 &    -0.0250 \\ 
    0.1000 &     0.4116 &     0.0248 &     0.0456 &     0.0247 &    -0.0644 &    -0.0250 \\ 
    0.2236 &     0.9638 &     0.0248 &     0.0787 &     0.0247 &    -0.0842 &    -0.0252 \\ 
    0.3162 &     1.3863 &     0.0250 &     0.1050 &     0.0247 &    -0.0952 &    -0.0253 \\ 
    0.5000 &     2.2073 &     0.0255 &     0.1669 &     0.0252 &    -0.0947 &    -0.0257 \\ 
    0.7071 &     3.1175 &     0.0264 &     0.2246 &     0.0261 &    -0.1157 &    -0.0271 \\ 
    1.0000 &     4.3219 &     0.0283 &     0.3425 &     0.0279 &    -0.1414 &    -0.0292 \\ 
    1.2247 &     5.1857 &     0.0303 &     0.4465 &     0.0296 &    -0.1464 &    -0.0315 \\ 
    1.4142 &     5.8561 &     0.0322 &     0.5769 &     0.0312 &    -0.1630 &    -0.0328 \\ 
    1.7321 &     6.9098 &     0.0361 &     0.7642 &     0.0355 &    -0.2293 &    -0.0366 \\ 
    2.0000 &     7.6365 &     0.0400 &     0.9622 &     0.0390 &    -0.2814 &    -0.0397 \\ 
    2.2361 &     8.1753 &     0.0436 &     1.1843 &     0.0420 &    -0.3371 &    -0.0432 \\ 
    2.4495 &     8.5577 &     0.0470 &     1.4165 &     0.0448 &    -0.3923 &    -0.0470 \\ 
    2.6458 &     8.8069 &     0.0502 &     1.6450 &     0.0473 &    -0.4364 &    -0.0497 \\ 
    2.8284 &     8.9725 &     0.0530 &     1.8567 &     0.0495 &    -0.4576 &    -0.0524 \\ 
    3.0000 &     9.1051 &     0.0556 &     2.1063 &     0.0518 &    -0.4779 &    -0.0548 \\ 
    3.1623 &     9.1304 &     0.0581 &     2.3100 &     0.0543 &    -0.4779 &    -0.0568 \\ 
    3.8730 &     8.9653 &     0.0685 &     3.1310 &     0.0632 &    -0.3378 &    -0.0664 \\ 
    4.4721 &     8.3308 &     0.0742 &     3.5479 &     0.0689 &    -0.1895 &    -0.0719 \\ 
    5.0000 &     7.5635 &     0.0773 &     3.7221 &     0.0726 &    -0.0771 &    -0.0755 \\ 
    5.4772 &     6.8348 &     0.0787 &     3.7424 &     0.0750 &     0.0275 &    -0.0779 \\ 
    5.9161 &     6.1931 &     0.0790 &     3.5938 &     0.0769 &     0.1044 &    -0.0800 \\ 
    6.3246 &     5.6684 &     0.0784 &     3.3780 &     0.0781 &     0.1714 &    -0.0810 \\ 
    6.7082 &     5.2233 &     0.0778 &     3.0897 &     0.0794 &     0.2627 &    -0.0820 \\ 
    7.0711 &     4.8000 &     0.0768 &     2.8576 &     0.0796 &     0.3891 &    -0.0841 \\ 
    7.4162 &     4.4129 &     0.0760 &     2.6405 &     0.0799 &     0.4688 &    -0.0841 \\ 
    7.7460 &     4.1362 &     0.0753 &     2.4296 &     0.0801 &     0.5019 &    -0.0835 \\ 
    8.0623 &     3.8792 &     0.0743 &     2.2182 &     0.0803 &     0.5101 &    -0.0829 \\ 
    8.3666 &     3.6840 &     0.0734 &     2.0377 &     0.0799 &     0.5102 &    -0.0823 \\ 
    8.6603 &     3.5341 &     0.0725 &     1.9025 &     0.0793 &     0.4961 &    -0.0821 \\ 
    8.9443 &     3.3927 &     0.0717 &     1.7905 &     0.0788 &     0.4761 &    -0.0821 \\ 
    9.2195 &     3.2770 &     0.0713 &     1.6856 &     0.0783 &     0.4460 &    -0.0818 \\ 
    9.4868 &     3.1878 &     0.0706 &     1.5982 &     0.0779 &     0.4268 &    -0.0818 \\ 
    9.7468 &     3.0968 &     0.0695 &     1.5040 &     0.0776 &     0.3909 &    -0.0820 \\ 
   10.0000 &     3.0169 &     0.0687 &     1.4089 &     0.0771 &     0.3398 &    -0.0816 \\ 
\hline
\hline
 \end{tabular}}
\caption{Force data for $\Gamma = 10$, $\beta = 10$, $\theta = 22.5^\circ$.}
\label{tab:G10_b10_t22p5}
\end{subtable}%
\begin{subtable}{0.5\textwidth} \centering
{\begin{tabular}{ccccccc}
\hline
\hline
$V$ & $-F_v$ & $\sigma_{F_v}$ & $F_\times$ & $\sigma_{F_\times}$ & $F_n$ & $\sigma_{F_n}$ \\
\hline
    0.0000 &    -0.9096 &     0.1972 &    -0.3797 &     0.1878 &     0.2402 &    -0.1901 \\ 
    0.0707 &     0.1144 &     0.1966 &    -0.3298 &     0.1867 &     0.2042 &    -0.1886 \\ 
    0.1000 &     0.5626 &     0.1967 &    -0.2956 &     0.1871 &     0.1928 &    -0.1890 \\ 
    0.2236 &     2.4385 &     0.1970 &    -0.2143 &     0.1855 &     0.1124 &    -0.1904 \\ 
    0.3162 &     3.8855 &     0.1976 &    -0.1759 &     0.1878 &     0.0949 &    -0.1886 \\ 
    0.5000 &     6.5994 &     0.1975 &    -0.0458 &     0.1860 &     0.0271 &    -0.1884 \\ 
    0.7071 &     9.7673 &     0.1978 &     0.1091 &     0.1868 &    -0.0943 &    -0.1888 \\ 
    1.0000 &    14.2137 &     0.1997 &     0.2103 &     0.1878 &    -0.1396 &    -0.1890 \\ 
    1.2247 &    17.6142 &     0.2014 &     0.3063 &     0.1898 &    -0.2346 &    -0.1923 \\ 
    1.4142 &    20.4609 &     0.2019 &     0.5022 &     0.1899 &    -0.2068 &    -0.1918 \\ 
    1.7321 &    25.2549 &     0.2037 &     0.8093 &     0.1907 &    -0.4001 &    -0.1968 \\ 
    2.0000 &    29.2578 &     0.2060 &     1.1644 &     0.1948 &    -0.5176 &    -0.1959 \\ 
    2.2361 &    32.6893 &     0.2090 &     1.4633 &     0.1981 &    -0.6124 &    -0.1997 \\ 
    2.4495 &    35.8198 &     0.2128 &     1.7379 &     0.2015 &    -0.7758 &    -0.2031 \\ 
    2.6458 &    38.4964 &     0.2165 &     2.0639 &     0.2059 &    -0.8592 &    -0.2078 \\ 
    2.8284 &    41.0747 &     0.2185 &     2.3632 &     0.2085 &    -0.9055 &    -0.2102 \\ 
    3.0000 &    43.3459 &     0.2211 &     2.6358 &     0.2143 &    -1.0278 &    -0.2177 \\ 
    3.1623 &    45.4009 &     0.2241 &     2.9730 &     0.2186 &    -1.1051 &    -0.2218 \\ 
    3.8730 &    54.6087 &     0.2369 &     4.4749 &     0.2292 &    -1.0785 &    -0.2295 \\ 
    4.4721 &    61.8219 &     0.2536 &     5.8976 &     0.2461 &    -1.6165 &    -0.2501 \\ 
    5.0000 &    68.0601 &     0.2666 &     7.3083 &     0.2536 &    -1.7616 &    -0.2559 \\ 
    5.4772 &    73.5036 &     0.2827 &     8.4863 &     0.2727 &    -2.0492 &    -0.2730 \\ 
    5.9161 &    78.1633 &     0.3012 &     9.5235 &     0.2877 &    -2.6086 &    -0.2947 \\ 
    6.3246 &    82.0425 &     0.3204 &    10.4761 &     0.3104 &    -2.9236 &    -0.3158 \\ 
    6.7082 &    85.0063 &     0.3414 &    11.4316 &     0.3289 &    -3.2169 &    -0.3378 \\ 
    7.7460 &    90.7990 &     0.4009 &    14.7987 &     0.3806 &    -4.7714 &    -0.3959 \\ 
    8.0623 &    91.8949 &     0.4196 &    15.9427 &     0.3933 &    -5.0937 &    -0.4108 \\ 
    8.3666 &    93.0761 &     0.4378 &    16.9343 &     0.4089 &    -5.3090 &    -0.4205 \\ 
    8.6603 &    93.8710 &     0.4547 &    18.1718 &     0.4186 &    -5.5267 &    -0.4394 \\ 
    8.9443 &    94.5805 &     0.4711 &    19.2223 &     0.4320 &    -5.7177 &    -0.4569 \\ 
    9.2195 &    95.0078 &     0.4876 &    20.1927 &     0.4456 &    -5.8926 &    -0.4733 \\ 
    9.4868 &    95.1345 &     0.5046 &    21.3494 &     0.4619 &    -6.0214 &    -0.4871 \\ 
    9.7468 &    95.4150 &     0.5197 &    22.3609 &     0.4732 &    -6.2370 &    -0.4961 \\ 
   10.0000 &    95.6099 &     0.5333 &    23.3016 &     0.4874 &    -6.5115 &    -0.5114 \\ 
   12.2474 &    91.1814 &     0.6401 &    32.6715 &     0.5784 &    -5.9760 &    -0.6188 \\ 
   14.1421 &    83.2466 &     0.6957 &    39.4653 &     0.6292 &    -3.2501 &    -0.6823 \\ 
   15.8114 &    74.1527 &     0.7262 &    41.4753 &     0.6846 &     0.4431 &    -0.7383 \\ 
   17.3205 &    65.1708 &     0.7418 &    42.1259 &     0.7164 &     4.4320 &    -0.7665 \\ 
   18.7083 &    57.4941 &     0.7393 &    40.7461 &     0.7364 &     7.2370 &    -0.7804 \\ 
   20.0000 &    51.2761 &     0.7337 &    38.8987 &     0.7374 &     9.2047 &    -0.7786 \\

\hline
\hline
 \end{tabular}}
\caption{Force data for $\Gamma = 100$, $\beta = 10$, $\theta = 22.5^\circ$.}
\label{tab:G100_b10_t22p5}
\end{subtable}
\end{table}